\documentclass{aa}  
\usepackage{lmodern}
\usepackage{float}

\usepackage[switch]{lineno}

\usepackage{graphicx}
\usepackage{txfonts}
\begin{document}

   \title{Probing the star formation main sequence down to 10$^{7}\, M_\odot$ at $1 < z < 9$}

   \author{Rosa M. Mérida
          \inst{1}
          \and
          Marcin Sawicki\inst{1} \and Kartheik G. Iyer\inst{2} \and Ga\"el Noirot\inst{3} \and Chris J. Willott\inst{4} \and Maru\v{s}a Brada\v{c}\inst{5,6} \and\\ Guillaume Desprez\inst{7} \and Nicholas S. Martis\inst{5} \and Adam Muzzin\inst{8} \and Gregor Rihtar\v{s}i\v{c}\inst{5} \and Ghassan T. E. Sarrouh\inst{8} \and Jeremy Favaro\inst{1} \and Gaia Gaspar\inst{1,9} \and Anishya Harshan\inst{5} \and Jon Jude\v{z}\inst{5}
          }

   \institute{
                Institute for Computational Astrophysics and Department of Astronomy and Physics, Saint Mary's University, 923 Robie Street, Halifax, NS B3H 3C3, Canada
              \email{Rosa.MeridaGonzalez@smu.ca}
         \and
             Columbia Astrophysics Laboratory, Columbia University, 550 West 120th Street, New York, NY 10027, USA
        \and
            Space Telescope Science Institute, 3700 San Martin Drive, Baltimore, Maryland 21218, USA
        \and
              National Research Council of Canada, Herzberg Astronomy \& Astrophysics Research Centre, 5071 West Saanich Road, Victoria, BC, V9E 2E7, Canada
        \and
            Faculty of Mathematics and Physics, Jadranska ulica 19, SI-1000 Ljubljana, Slovenia
        \and
            Department of Physics and Astronomy, University of California Davis, 1 Shields Avenue, Davis, CA 95616, USA
        \and
            Kapteyn Astronomical Institute, University of Groningen, P.O. Box 800, 9700AV Groningen, The Netherlands
        \and
            Department of Physics and Astronomy, York University, 4700 Keele St. Toronto, Ontario, M3J 1P3, Canada
        \and
            Observatorio Astronómico de Córdoba, Universidad Nacional de Córdoba, Laprida 854, X5000, Córdoba, Argentina}
   \date{Received September 15, 1996; accepted March 16, 1997}

  \abstract
   {The main sequence of star-forming galaxies (SFGMS or MS) is a fundamental scaling relation that provides a global framework for studying galaxy formation and evolution, as well as an insight into the complex star formation histories (SFHs) of individual galaxies. In this work, we combine large-area pre-JWST surveys (COSMOS2020, CANDELS), which probe high-$M_\star$ sources (${>10^9\,M_\odot}$), with SHARDS/CANDELS FAINT and JWST data from CANUCS, CEERS, JADES, and UNCOVER, to obtain a high-$z$, star formation rate (SFR) and stellar mass ($M_\star$) complete sample spanning both high- and low-$M_\star$ regimes. Completeness in both $M_\star$ and the SFR is key to avoiding biases introduced by low-mass, highly star-forming objects. Our combined dataset is 80\% complete down to $10^{7.6}\,M_\odot$ at $z\sim1$ ($10^{8.8}\,M_\odot$ at $z\sim9$). The overall intrinsic MS slope (based on the SFR$_{100}$ and $M_\star$ derived with \texttt{Dense Basis} and nonparametric SFHs) shows little evolution up to $z\sim5$, with values $\sim0.7 - 0.8$. The slope in the low-$M_\star$ regime becomes steeper than that in the high-$M_\star$ end at least up to $z\sim5$, but the strength of this change is highly dependent on the assumptions made on the symmetry of the uncertainties in $M_\star$ and SFR. If real, the steepening suggests reduced star formation efficiency or declining gas content with decreasing $M_\star$. The transition between the low-$M_\star$ regime and the canonical MS occurs around $10^{9.5}\,M_\odot$, independent of $z$. This critical value may coincide with the assembly of galaxies' disks, which can provide a mechanism for self-regulation that stabilizes them against feedback. The intrinsic scatter is compatible with canonical estimates, also at low-$M_\star$, ranging from $0.2-0.3$ dex. This is indicative of rapid variations in star formation being averaged out over $\lesssim100$~Myr.}

   \keywords{Galaxies: photometry -- star formation -- high-redshift --
                evolution
               }

   \maketitle
%
%-------------------------------------------------------------------

\section{Introduction}
\label{sec:intro}

Scaling relations manifest the combined action of the diverse and complex physical processes undergone by galaxies. They serve as benchmarks for any theoretical model exploring galaxy formation and evolution.
These relations allow us to look past the plethora of local-scale phenomena, predict future evolutionary phases, and provide a cohesive global picture built from direct observables (e.g., the Schmidt-Kennicutt law of star formation; \citealt{Schmidt1959}, \citealt{Kennicutt1989}).

Despite the power of these tools, the overall picture of the formation and evolution of galaxies is still incomplete. Before the arrival of JWST, observing the faint Universe at high $z$ was a challenging goal. Our understanding of the early epochs was based on observations of bright objects (Malmquist bias; \citealt{Malmquist1922}). However, these bright
sources may not exhibit or be primarily governed by the same processes that dominate
fainter galaxies. 
Some of these faint objects correspond to massive systems whose emission remains undetected in the rest-frame optical, with more prominent spectral energy distributions (SEDs) at longer wavelengths (e.g., star-forming galaxies, SFGs, with high dust content; see \citealt{Alcalde-Pampliega2019}, \citealt{Zhou2020}). However, a vast
number of objects that lie beyond the sensitivities of pre-JWST optical and near-infrared (NIR) capabilities are actually low-mass galaxies (e.g., \citealt{Austin2023}; \citealt{Merida2023}, M+23 hereafter; \citealt{Strait2023}; \citealt{Asada2023,Asada2024}; \citealt{Looser2024}). 

The mass-completeness limits achieved thanks to the \textit{Hubble} Space Telescope (HST) sensitivity allowed us to reach down to $10^{8.6}\, M_\odot$ at $z = 1$ ($10^{10} \,M_\odot$ at $z = 8$) (\citealt{Barro2019}, B+19 hereafter).
Some strategies that enabled the recovery of lower-luminosity (lower-mass) galaxies at high-$z$ in the HST era involved exploiting gravitational lensing (e.g., \citealt{Caputi2021}, \citealt{Sun2021}); absorption systems (e.g., \citealt{Lofthouse2020}, \citealt{Joshi2025}); emission lines (e.g., {\citealt{Maseda2014}, \citealt{Atek2022}}); or stacking techniques (e.g., M+23, \citealt{Merida2024}, \citealt{Wang2024}). Nowadays, these limits are easy to push thanks to the depth and sensitivity of new-generation telescopes.

However, even if we can detect these low-mass systems, it is important to note that the calibrations and conversions used to compute integrated quantities were derived from the low-$z$ picture. Looking for analogs of low-$z$ low-mass galaxies, such as extremely metal-poor galaxies, blue compact dwarfs, or damped Lyman-$\alpha$ systems (e.g., \citealt{Wolfe2005}, \citealt{Cairos2010}, \citealt{Lorenzo2022}), was key. Now, we can start studying the interstellar medium (ISM) of low-mass high-$z$ galaxies themselves (e.g., \citealt{Lin2023}, \citealt{Harshan2024}, \citealt{Looser2024b}, \citealt{Deugenio2025}), which provides the means to improve calibrations.

A powerful scaling relation that was often extrapolated to lower stellar masses ($M_\star$) is the so-called main sequence of star-forming galaxies (SFGMS or MS; e.g., \citealt{Daddi2007}, \citealt{Noeske2007}, \citealt{Elbaz2011}, \citealt{Sawicki2012}, \citealt{Whitaker2012}, \citealt{Lee2015}, \citealt{Santini2017}, \citealt{Barro2019}, M+23). It describes the tight correlation observed between $M_\star$ and the star formation rate (SFR), seen to persist from the times of the local Universe back to the Epoch of Reionization.
This correlation is typically parameterized using a power law (i.e., log SFR = log~$M_\star\times\beta$ + $\alpha$, where $\alpha$ is the normalization and $\beta$ the slope), although the exact shape of the relation and its evolution is still a matter of debate \citep{Popesso2023}.

Different SFG selections, normally based on color-color diagrams (e.g., $UVJ$; \citealt{Williams2009}, \citealt{Whitaker2011}; $NUVrK$, \citealt{Arnouts2007}, \citealt{Noirot2022}) or specific SFR (sSFR) cuts, can lead to different results \citep{Antwi-Danso2023}. Additionally, rest-frame $J$ fluxes at $z > 3$ were less accurate in past studies because they were extrapolated; at these
$z$, this information was derived from the \textit{Spitzer} Infrared Array Camera
bands, for which data are usually shallower. Different SFR indicators can also affect results, especially when dust obscuration corrections are uncertain.

Despite these possible sources of discrepancy, there is some consensus on the constraints of the MS parameters; namely, the normalization, slope, and scatter ($\epsilon$).
The normalization of the MS has been shown to increase smoothly with $z$, likely on timescales that depend on a galaxy's mass
(e.g., \citealt{Speagle2014}, \citealt{Pearson2018}, \citealt{Popesso2023}). The evolution of this parameter indicates that star formation activity was more intense in earlier epochs.

The MS slope is related to the star formation efficiency (SFE) and the gas content in galaxies. When gas is unavailable or not efficiently consumed, the MS may take on a steeper slope. Many authors claim that there is little to no evolution in the slope with $z$, and that its value is close to unity (e.g., \citealt{Dunne2009},
\citealt{Whitaker2012}, \citealt{Kashino2013}, \citealt{Santini2017}), although values range from $0.6 - 1.0$ in the literature (\citealt{Speagle2014} and references therein). This wide range of slopes is partially due to the lower and upper mass limits considered, which vary from study to study (see M+23).
Some works suggest that the
MS slope flattens at high $M_\star$, above a turnover point that is $z$-dependent
($\sim10^{10.2-10.5}\, M_\odot$; e.g., \citealt{Whitaker2014}, \citealt{Lee2015}, \citealt{Schreiber2015},
\citealt{Tomczak2016}). They propose a scenario in which quenching of star formation in the disks of massive galaxies could be driven by the interplay between bulge growth \citep{Abramson2014} and active galactic nucleus (AGN) feedback \citep{Leslie2016}.

The MS scatter, which ranges from $0.2-0.3$ dex (see \citealt{Speagle2014} and references therein), is framed within the self-regulation evolution model (see \citealt{Lilly2013}, \citealt{Tacchella2016}). The small scatter would reflect the interplay between gas accretion and feedback processes that make galaxies' SFRs fluctuate on short timescales along the MS. However, this scatter would be purely statistical if the MS were instead showing the “population average” at a certain age (i.e., if the MS were not ergodic; \citealt{Smith2024}). In that case, galaxies would fluctuate around different median trajectories, related to their halo mass and halo formation times (see \citealt{Abramson2016}, \citealt{Matthee2019}). On the other hand, a tighter scatter can also be induced by assuming too smooth star formation histories (SFHs) in SED modeling \citep{Bohorquez2025}.

According to the self-regulated evolution model, galaxies follow smooth and steady star formation trajectories during most of their lives. However, high-$z$ galaxies, and especially low-$M_\star$ galaxies, undergo many periods of bursting and latency, with more stochastic SFHs (see \citealt{Asada2023,Asada2024}, \citealt{Looser2024}, \citealt{Bohorquez2025}, \citealt{Cole2025}, \citealt{Mintz2025}, \citealt{Simmonds2025} for an observational perspective; \citealt{Stinson2007}, \citealt{Dominguez2015}, \citealt{Flores2021}, \citealt{Gelli2023}, \citealt{Sun2023} for a simulation point of view; and \citealt{Broussard2019}, \citealt{Emami2019}, and \citealt{Gelli2025} for a hybrid perspective). Such differences may be reflected in the behavior of the MS at lower $M_\star$. A more stochastic SFH should yield a larger MS scatter, whereas a change in the SFE would affect the MS slope, which would likely show an inflection point similar to the one reported at high $M_\star$ (see \citealt{Ma2018}, M+23, \citealt{Ciesla2024}).

Several works attempted to push the MS down to $10^8\,M_\odot$ at $0<z<3$, finding a slope and scatter compatible with the values retrieved at higher $M_\star$ (e.g., \citealt{Reddy2012}, \citealt{Sawicki2012}, \citealt{Santini2017}, \citealt{Boogaard2018}). In M+23 we probed similar $M_\star$ with a much larger dataset, revealing hints of a possible steepening of the MS and a scatter compatible with $\sim0.3$~dex.
On the other hand, studies such as \citet[see also \citealt{Rinaldi2025}]{Rinaldi2022} and \cite{Clarke2024} find flatter slopes and a larger scatter in the low-$M_\star$ regime (see also \citealt{Cole2025}). However, this behavior can be partially due to selection effects if $M_\star$ and SFR completeness limits are overestimated (Mérida et al. in prep). 

The behavior of the MS at lower $M_\star$ thus remains unclear, especially as we move to increasingly higher $z$. Moreover, whether the MS slope increases or decreases, any intrinsic change in the MS should be considered with caution, given the possible lack of symmetry in the uncertainties and the typical low signal-to-noise (S/N) of photometric points at these redshifts and $M_\star$, which can make the codes overfit the SEDs and be biased toward the values of the priors.

We now turn to the specific outline of this work, which adopts a wedding-cake approach to studying the MS.

When fitting the MS, or any scaling relation, different challenges can arise. Systematic biases can appear when different codes and configurations are considered, which can lead to wrong interpretations of the evolution of MS parameters. Such offsets can also arise when combining data from different instruments. For instance, JWST is more sensitive than previous instruments to low-surface-brightness, low-SFR galaxies, which can lead to an MS offset with respect to the known normalization. Lastly, incompleteness toward lower $M_\star$ and SFR values can also induce artificial changes in the MS parameters. 

\begin{figure}[htp]
    \centering
    \includegraphics[width=\linewidth]{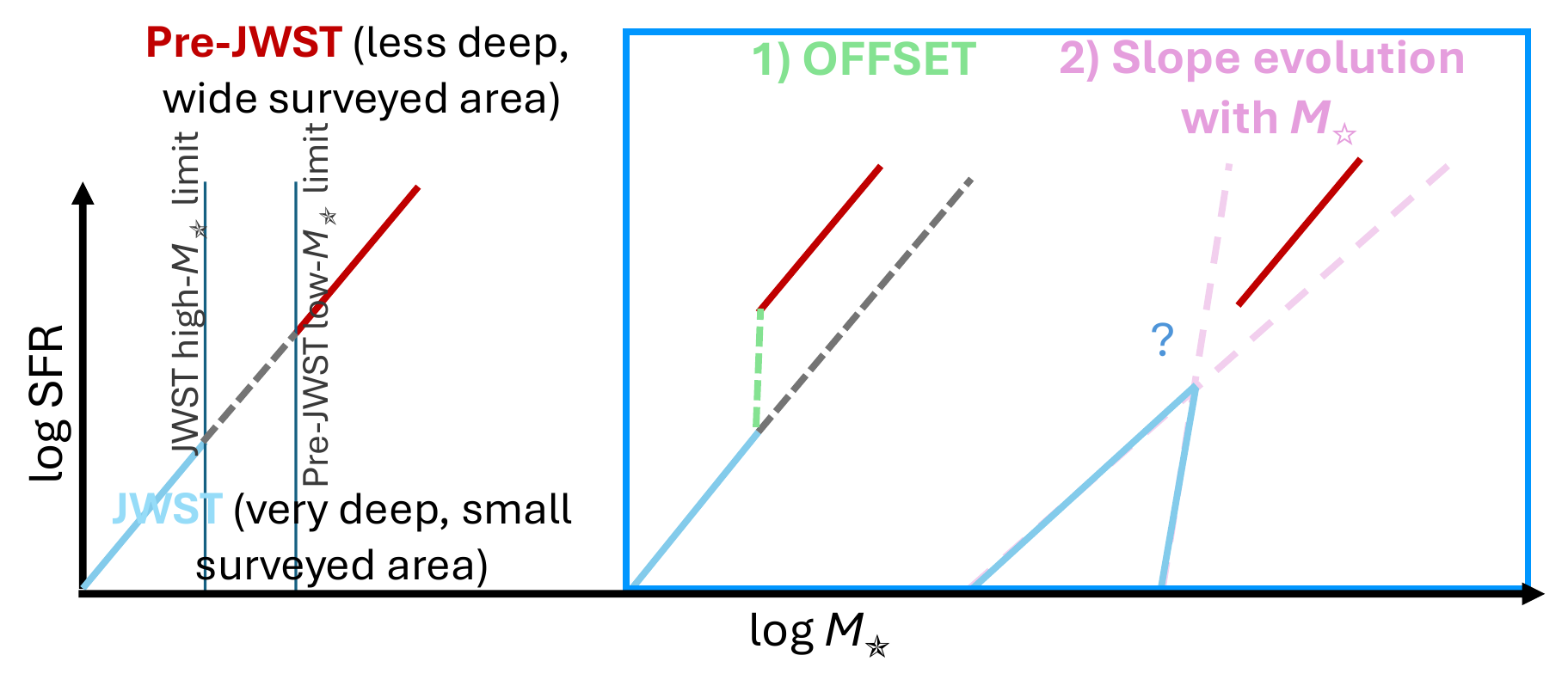}
    \caption{Cartoon depicting different challenges that can arise when trying to fit the MS with no continuous and complete coverage. Current JWST surveys are very deep, but limited in size, selecting mainly low-mass galaxies. Pre-JWST surveys are not deep enough to reach such low $M_\star$. Offsets and/or changes in the slopes can make the connection between the two regimes hard to quantify if we do not consider the ensemble of both pre-JWST and JWST surveys.}
    \label{fig:ms_drawing}
\end{figure}

The first two challenges can be addressed by applying a homogeneous method to measure galaxy properties to a sample that continuously probes the $M_\star - $SFR plane. 
Such a global picture provides the means to understand the physical inflection points that reflect the boundaries of the different regimes of the MS. This is key when trying to determine when the bursty stochastic star formation activity gives way to the smooth accretion mode. 
Figure~\ref{fig:ms_drawing} schematically depicts the difficulty of understanding the MS evolution with $M_\star$ when not considering the ensemble of both low- and high-$M_\star$ galaxies.

In this work we explore the MS evolution between ${1<z<9}$, spanning an unprecedentedly large range of $M_\star$ and SFR, enabled by an SFR and mass-complete sample that includes sources at the low-$M_\star$ end ($\sim10^7\,M_\odot$).
This robust and cohesive dataset comprises large-area pre-JWST surveys, such as the Cosmic Assembly Near-infrared Deep Extragalactic Legacy Survey (CANDELS; \citealt{Grogin2011}, \citealt{Koekemoer2011}) and COSMOS2020 \citep{Weaver2022}, and current JWST surveys. Pre-JWST surveys allow us to probe high-$M_\star$ ($>10^9 \, M_\odot$), while JWST data are crucial for extending completeness to lower $M_\star$. As such, this sample allows us to probe the MS across all $M_\star$. Throughout this work we assumed $\Omega_\mathrm{M,0}=0.3$, $\Omega_{\Lambda,0}=0.7$, H$_0=70$ km s$^{-1}$ Mpc$^{-1}$, and a \citet{Chabrier2003} initial mass function, and used AB magnitudes \citep{Oke1983}.

\section{Data and sample selection}
\label{sec:data_sample}

In this work we aim to study the evolution of the MS in the largest possible and most complete $M_\star$ and SFR range. For that purpose, we combined large-area pre-JWST surveys, which allow us to investigate the high-$M_\star$ regime ($>10^9\,M_\odot$), with JWST surveys that enable the completeness to be pushed toward lower $M_\star$. The pre-JWST surveys selected in this work were CANDELS and COSMOS2020 (which sum up $>1$ million galaxies), and SHARDS/CANDELS FAINT (M+23). The latter is based on image stacking and allowed us to push the limits of the former CANDELS catalog in GOODS-N toward lower $M_\star$ values. This dataset provides an independent check on the trends observed at the low-$M_\star$ end by JWST.

We selected JWST galaxies from the Cosmic Evolution Early Release Science Survey (CEERS;
ERS-1345, \citealt{Finkelstein2022}, \citealt{Bagley2023}); The \textit{\textit{James Webb}} Space Telescope Advanced Deep Extragalactic Survey (JADES; GTO-1180, 1181, 1210, 1286, GO-1895, 1963, \citealt{Eisenstein2023}); The CAnadian NIRISS Unbiased Cluster/Technicolor Survey (CANUCS/Technicolor; GTO-1208, GO-3362, \citealt{Willott2022}, \citealt{Sarrouh2025}); and The Ultradeep NIRSPec and NIRCam Observations before the
Epoch of Reionization survey (UNCOVER; GO-2561, \citealt{Bezanson2024}). A more detailed description of each of these datasets can be found in Appendix~\ref{app:surveys}. The final number of objects from each survey included in this work, together with some average properties, can be consulted in Table~\ref{tab:data}.

\begin{table}[htp]
\setlength{\tabcolsep}{1.8pt}
    \centering
    \small
    \caption{Surveys used in this work, ordered by the covered area.}
    \begin{tabular}{c|c|c|c|c|c|c}
    Survey & Area& \# Objects& $\overline{z}$ &$\overline{\mathrm{log} M_\star}$&$\overline{\mathrm{log SFR}_{100}}$&$A_v$\\
    & arcmin$^2$ & & &$M_\odot$ & $M_\odot$ yr$^{-1}$&mag\\
    \hline\hline
    COSMOS2020 & 7,200 &551,824 & 2.3& 9.2& 0.5&0.4\\ \hline
    CANDELS & 941 & 85,636& 2.1& 8.8& 0.2&0.4\\
    \hline
    JADES & 175 & 43,303 &3.1& 7.9& $-0.7$&0.3\\ \hline
    SHARDS/CANDELS& 122 &12,439 & 2.1& 7.6& $-0.6$&0.2\\
     FAINT &  & & & & &\\ \hline
    CEERS & 100 & 9,933 &1.8& 8.5& $-0.4$&0.2\\ \hline
    CANUCS & 100 &27,827 & 2.7& 8.1& $-0.4$&0.2\\ \hline
    UNCOVER & 56 & 14,234 &2.7& 8.0& $-0.6$&0.3\\ \hline
    \end{tabular}
    \label{tab:data}
\tablefoot{The total surveyed area obtained by combining the coverage of the different catalogs is $\sim2.2$ deg. The numbers listed in column 2 correspond to the final samples, obtained after applying the screenings described in Sect. \ref{sec:data_sample} and \ref{sec:properties}. The average $z$, log $M_\star,$ log SFR$_{100}$, and $A_v$, listed in columns $4-7$, were estimated based on the galaxies that entered this selection.}
\end{table}

To construct our parent sample, we applied several screenings to this dataset, as is outlined below.
Given that some of these surveys overlap, namely JADES and CANDELS/GOODS, CEERS and CANDELS/EGS, and COSMOS2020 and CANDELS/COSMOS, we first removed potential duplicate galaxies by looking for counterparts within 0.3\arcsec\, and keeping only the CANDELS sources. This guarantees a uniform treatment of overlaps, assigning them to the same parent sample, and weighting them equally in further analysis (see Sect.~\ref{sec:db}). SHARDS/CANDELS FAINT and CANDELS/GOODS-N also overlap and a similar process was already carried out in M+23. These duplicates constitute $\sim$~30\% of the CEERS sources and $\sim$~10\% of the JADES and COSMOS2020 galaxies, respectively. 

These preliminary catalogs were then trimmed to keep galaxies with useful measurements in enough bands, considering filters covering the $0.3 - 8$~$\mu$m wavelength range, through the \texttt{nbands} or \texttt{EAZY\_nfilt} flags if available, setting a $\geq8$ value. In addition to this cut, we required S/N~$\geq3$ in at least eight bands, which is important especially when the previous flags are not available. We used the \texttt{use$\_$phot} flag when available too. 

For the UNCOVER and CANUCS cluster field catalogs, we restricted the gravitational lensing magnification to ${1<\mu<10}$, correcting SFRs, $M_\star$, and their uncertainties based on the values provided by each catalog. 
The redshifts, also taken from the catalogs, were restricted to $1 \leq z \leq9$ in every survey. These cuts left us with a parent sample of 824,287 galaxies to be fit: 613,261 from COSMOS2020; 96,948 from CANDELS; 12,696 from SHARDS/CANDELS FAINT; 10,859 from CEERS; 43,853 from JADES; 31,800 from CANUCS; and 14,870 from UNCOVER. Note that this is not our final sample, as subsequent screenings are still required. These are detailed in Sect.~\ref{sec:db}.

\section{Physical properties}
\label{sec:properties}

\subsection{Obtaining homogeneous $M_\star$ and SFR measurements}
\label{sec:db}

The physical properties of each of the catalogs listed in Sect.~\ref{sec:data_sample} were originally computed using different fitting codes. CANDELS and SHARDS/CANDELS FAINT galaxies were fit using \texttt{EAzY} \citep{Brammer2008} and \texttt{pzeta} \citep{Perez-Gonzalez2005,Perez-Gonzalez2008} for the photo-$z$s and \texttt{Synthesizer} \citep{Perez-Gonzalez2005,Perez-Gonzalez2008} for the stellar properties. The COSMOS2020 catalog was fit using \texttt{EAzY} and \texttt{LePhare} \citep{Arnouts2011}. The CANUCS team used \texttt{EAzY}, \texttt{Bagpipes} \citep{Carnall2018}, and \texttt{Dense Basis} (\citealt{Iyer2017}, \citealt{Iyer2019}). The UNCOVER catalog is based on \texttt{EAzY} and \texttt{Prospector-$\beta$} \citep{Johnson2021}; the one based on JADES used \texttt{EAzY} supplemented by flexible stellar population synthesis (\texttt{FSPS}; \citealt{Conroy2010}) templates; and CEERS relied on \texttt{EAzY}, \texttt{Bagpipes}, and \texttt{Dense Basis}. 

Some of these codes did not consider the effects of outshining (\citealt{Narayanan2024}, \citealt{Whitler2023}), which can lead to systematic underestimates of the $M_\star$ \citep{Sorba2018}. The latter is tightly linked to the SFH prescription, since some of them did not allow a nonparametric SFH to be set. The assumption of the priors and SFHs is more essential than the selection of the code itself. Nonparametric models capture better the variety of physical events (e.g., bursts, rejuvenation, sudden quenching, or maximally old star formation) responsible for the complex SFHs of high-$z$ galaxies (\citealt{Simha2014}, \citealt{Leja2019}, \citealt{Bohorquez2025}).

\begin{figure*}[htp]
    \centering
    \includegraphics[width=\linewidth]{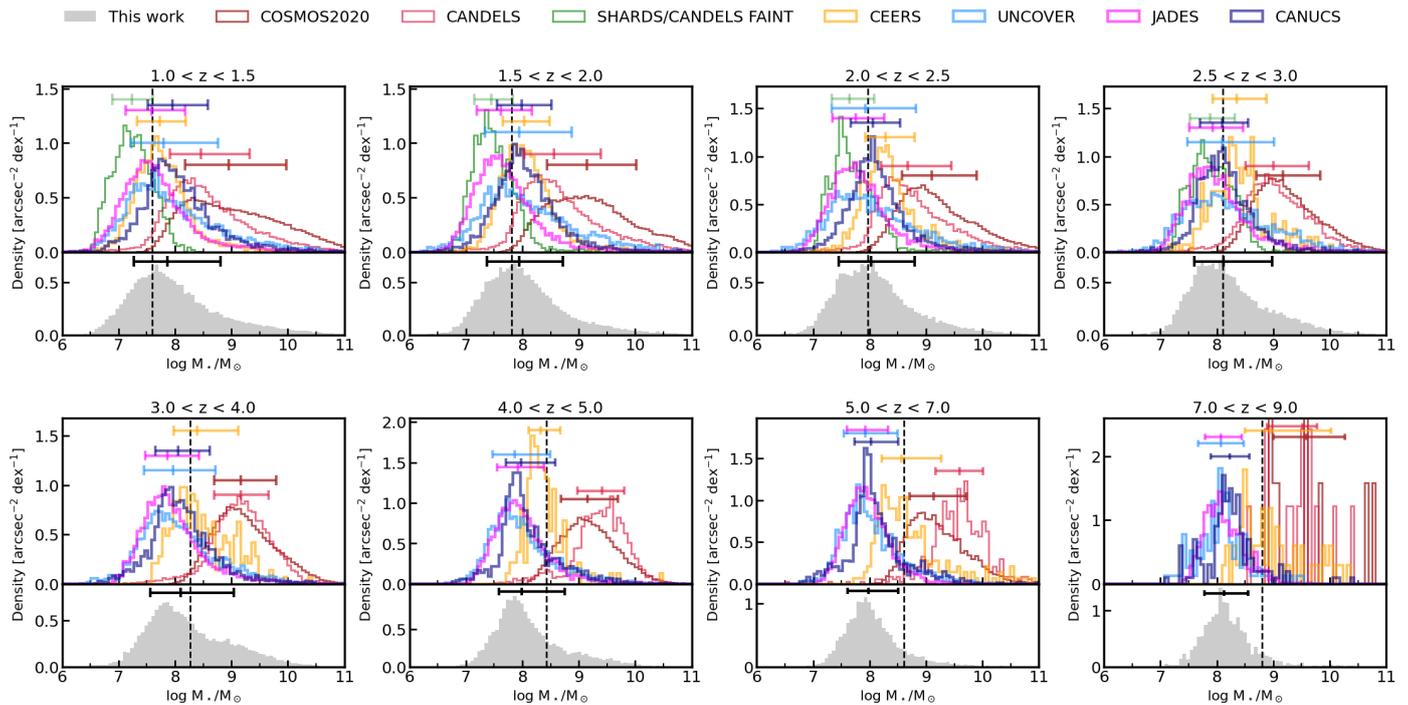}
    \caption{Stellar mass histograms in different redshift intervals showing the distribution of the CANDELS (red), COSMOS2020 (maroon), SHARDS/CANDELS FAINT (green), JWST/CEERS (yellow), JADES (fuchsia), CANUCS (navy), and UNCOVER (blue) surveys. Each histogram is normalized such that the total area under the curve equals 1. The solid gray histograms in the panels underneath show the distribution of the galaxies that entered our final selection, combining the previous surveys (see Sect.~\ref{sec:properties}). The 16th, 50th, and 84th percentiles for each distribution are displayed as segments, color-coded following the histograms (black for our final sample). Vertical dashed lines depict the 80\% mass-completeness limit of our sample (see Sect.~\ref{sec:completeness}).}
    \label{fig:mass}
\end{figure*}

To avoid possible systematic effects produced by combining results obtained from different configurations and codes (see Sect.~\ref{sec:intro}), we opted to refit all the galaxies within our sample using \texttt{Dense Basis} (\citealt{Iyer2017}, \citealt{Iyer2019}). 
This software uses nonparametric SFHs and incorporates stellar population synthesis models through the \texttt{FSPS} \texttt{Python} module. The code builds the SFHs based on the lookback times at which a galaxy assembles certain quantiles of its $M_\star$, using Gaussian processes to create smooth and flexible SFHs. It uses a flexible number of parameters, adjusted to extract the maximum amount of information from the SEDs being fit.

It is important to highlight that these are stellar fits only, and a possible AGN flux contribution is not considered. Recent studies have shown that $5 - 15$\% of galaxies at $z \sim 4 - 7$ host supermassive black holes (\citealt{Harikane2023}, \citealt{Maiolino2024}). Photometric data are sometimes insufficient to confirm or dismiss the presence of AGNs, and we warn the reader that our $M_\star$ and SFR estimates may be overestimated in some of our galaxies, especially at the high-$M_\star$ end.

We built our atlas, which is a grid made up of all the parameters' values, assuming wide ranges for the priors of the redshift, $M_\star$, and SFR, constraining the $z$ between ${0 < z < 12}$; $M_\star$ between \mbox{$5<\mathrm{log}\, M_\star/M_\odot<12$}; and sSFR between \mbox{$-12<\mathrm{log\, sSFR} [\mathrm{yr^{-1}}]<-7.5$}, setting a flat sSFR prior. We considered the default configuration for the metallicities and a \citet{Calzetti2000} attenuation law with a flat prior for the dust. When running the code, the $z$ was fixed to the values provided by the different catalogs ($z_{\mathrm{spec}}$ when available). The atlas comprises $\texttt{N$\_$pregrid} = 1,000,000$ templates, and we allow these to explore redshifts around $\pm0.1$ the redshift of the galaxy.

We used point-spread function (PSF) matched photometry measured in Kron apertures \citep{Kron1980} for the COSMOS2020, CEERS, UNCOVER, CANUCS, and JADES objects. 
CANDELS photometry was also measured in Kron apertures using \texttt{TFIT} (\citealt{Laidler2007}; see also B+19). \texttt{TFIT} is a template-fitting code that enables one to measure galaxy photometry using prior information about the position of the sources from high-resolution observations (in this case, $F160W$). It creates mock PSF-matched images, computed on an object-by-object basis by smoothing the
high-resolution cutouts using a convolution kernel. \texttt{TFIT} then fits iteratively for the photometry by comparing the real images to the models, accounting for possible contamination from nearby galaxies.
SHARDS/CANDELS FAINT considers distinct apertures and resolutions and accounts for these differences through a series of aperture and offset corrections that yield results consistent with PSF-matched photometry. 

The resulting \texttt{Dense Basis} catalogs were also screened to remove quiescent galaxies using the $U$, $V$, and $J$ colors when provided by the photometric catalogs, and an additional sSFR cut dependent on $z$ ($\mathrm{log\: sSFR}>-10$ yr$^{-1}$ at $1<z<2.5$; $\mathrm{log\: sSFR}>-9.75$ yr$^{-1}$ at $2.5<z<3$; $\mathrm{log\: sSFR}>-9.5$ yr$^{-1}$ at $3<z<4$; and $\mathrm{log\: sSFR}>-9$ yr$^{-1}$ at $z>4$). These sSFR cuts are based on a visual inspection of the $\mathrm{SFR}-M_\star$ plane \citep{Pearson2023}. The $UVJ$ cuts, which evolve with $z$, were estimated using the code presented in \citet[see also \citealt{Tan2024}]{Antwi-Danso2023}. These colors correspond to the best-fit SED colors, not corrected with the observed photometry, which can affect our selection (see \citealt{Noirot2022}). However, these cuts affect mainly galaxies from CANDELS and COSMOS2020 with $M_\star>10^9\,M_\odot$ (see red points in Fig.~\ref{fig:MS}).

\begin{figure*}[htp]
    \centering
    \includegraphics[width=\linewidth]{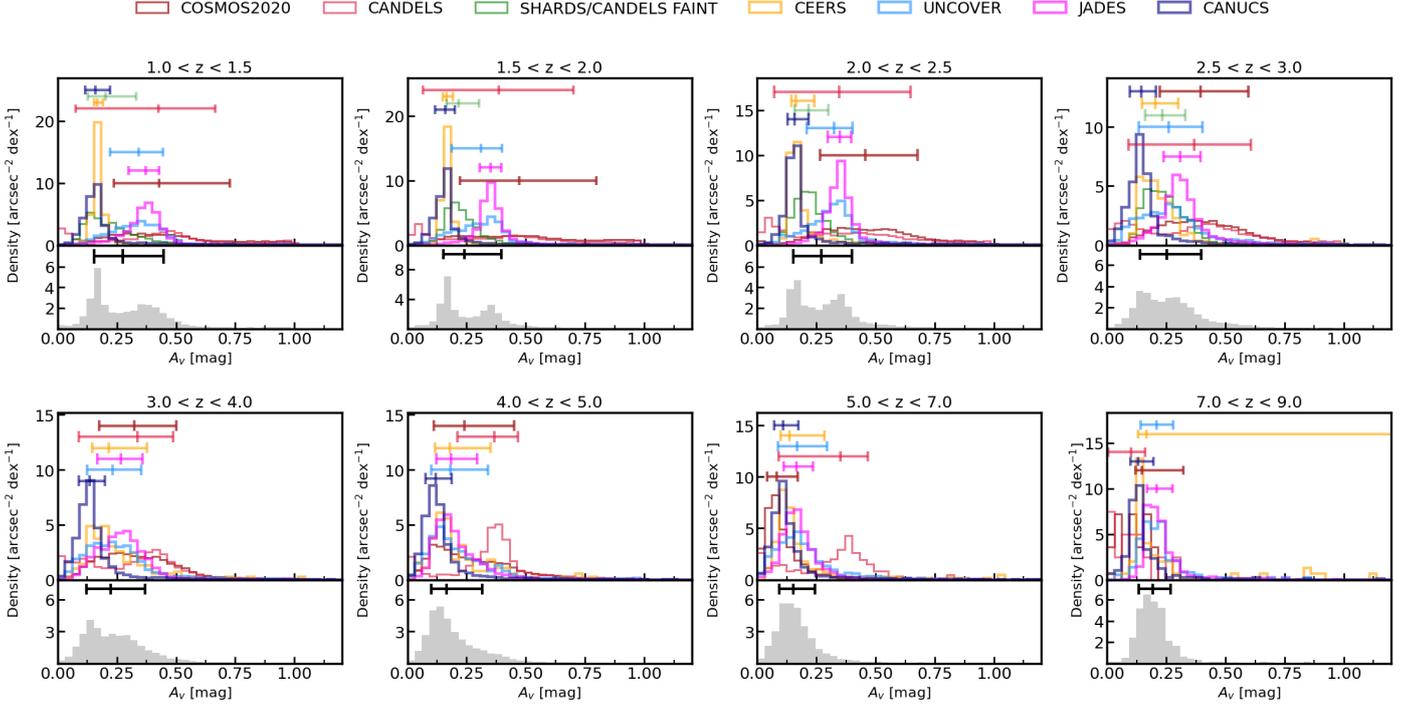}
    \caption{Attenuation histograms in different redshift intervals showing the distribution of the galaxies within the distinct surveys, following the same color code as in Fig.~\ref{fig:mass}.}
    \label{fig:Av}
\end{figure*}

\begin{figure*}[htp]
    \centering
    \includegraphics[width=\linewidth]{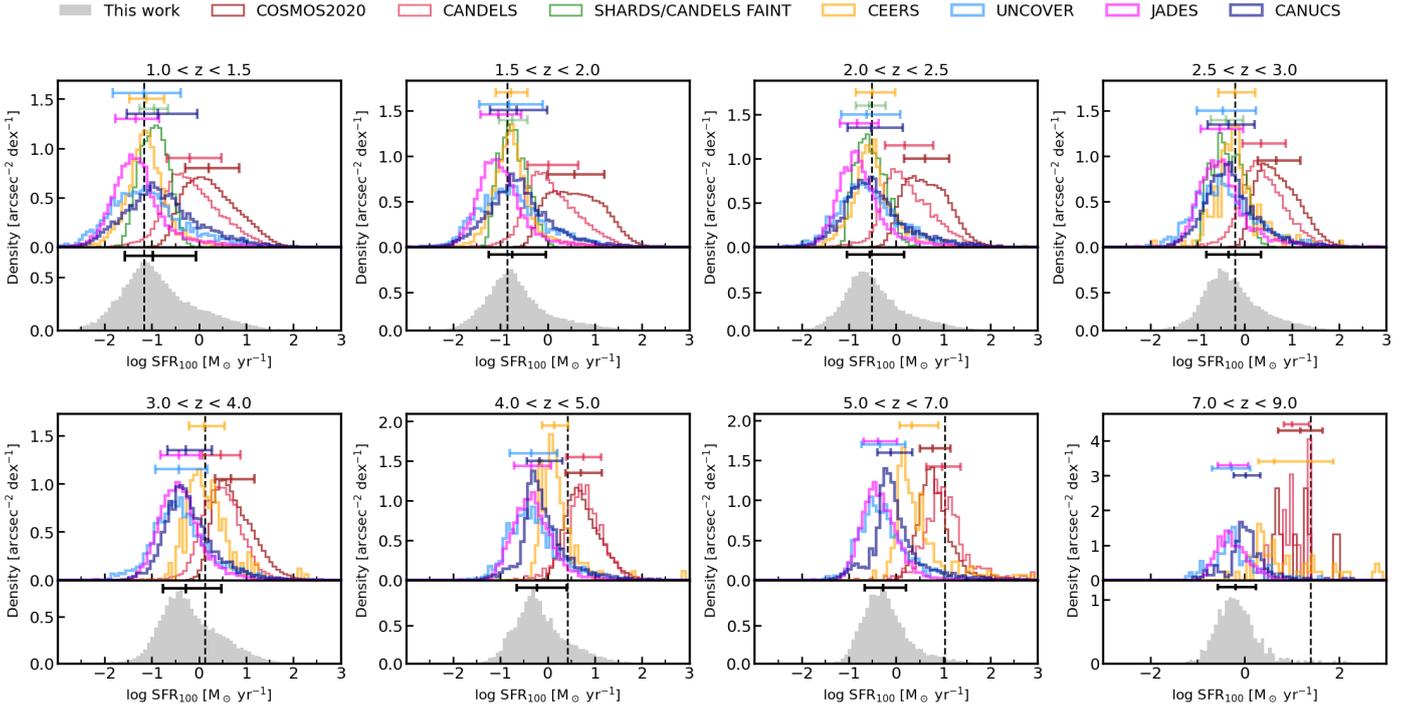}
    \caption{Star formation rate histograms in different redshift intervals showing the distribution of the galaxies within the distinct surveys, following the same color code as in Fig.~\ref{fig:mass}. Vertical dashed lines highlight our SFR limits (see Sect.~\ref{sec:completeness}).}
    \label{fig:sfr}
\end{figure*}

Moreover, we removed sources whose SFR or $M_\star$ 16th and 84th percentiles were not well constrained (i.e., were infinite or “not a number”, NaN). These cuts provided us with our final sample, made up of 755,196 galaxies. The contribution from each survey is listed in Table \ref{tab:data}.
Figures \ref{fig:mass}, \ref{fig:Av}, and \ref{fig:sfr} show the distribution in log $M_\star$, $A_v$, and log SFR$_{100}$ of the different subsamples, weighted by their surveyed area, and normalized such that the total area under the histograms is one. From now on, we shall use SFR to refer to SFR$_{100}$, or SFR$_{\mathrm{UV}}$, which probes the star formation within a 100 Myr timescale, being more robust against short-term variations of the star formation. A study of burstiness, using SFR$_{10}$, will be presented in Mérida et al. (in prep).

Figures~\ref{fig:mass}, \ref{fig:Av}, and \ref{fig:sfr} show that CANDELS and COSMOS2020 probe high-$M_\star$ galaxies, with, on average, higher SFR. JWST surveys and SHARDS/CANDELS FAINT probe lower $M_\star$ and SFRs. While there is a smooth transition between the mass regimes dominated by JWST and pre-JWST surveys at low $z$, there is gradually less overlap at increasingly high $z$. Wider JWST surveys are essential for populating the $10^{8.5-9.5} M_\odot$ regime at higher $z$, bridging the two $M_\star$ regimes. A step forward in this direction is, for instance, the MINERVA survey \citep{Muzzin2025}: an ongoing eight-filter medium-band Near Infrared Camera (NIRCam, \citealt{Rieke2023}) imaging survey in the primary HST/JWST fields covering $\sim548$~arcmin$^2$ of the sky.

In terms of extinction, CANDELS and COSMOS2020 display the highest values (as they also probe higher $M_\star$), although these are still small ($A_v<0.5$~mag). The average optical extinction of our final sample ranges from $0.2-0.3$~mag.

Given that each catalog covers a specific $M_\star$ and SFR range, with different depths and areas, assigning equal weights to the galaxies in these surveys would lead to skewed results. As a specific example, COSMOS2020 provides ${\sim600,000}$ sources, some of them low-mass ($10^{7-8}\,M_\odot$). However, the catalog is not mass-complete down to those values, and hence the low-mass population included in this survey is likely very bright and made up of starbursts. On the other hand, CANUCS covers a much smaller area and provides a lower number of sources. However, the survey is much deeper, allowing us to detect not only the brightest objects at the low-mass end but also the entire population of these galaxies. 

To address these differences, we selected galaxies based on the relative area covered by each survey with respect to the total area ($\sim2.2$ deg). We considered ${1\leq z<1.5}$, ${1.5\leq z<2}$, ${2\leq z<2.5}$, ${2.5\leq z<3}$, ${3\leq z<4}$, ${4\leq z<5}$, ${5\leq z<7}$, and ${7\leq z\leq 9}$ redshift intervals, dividing the log $M_\star$ range within each interval into 0.1 dex bins. For each bin, we randomly drew $N/k$ sources per catalog, $N$ being the number of galaxies in the bin and $k$ the relative area (i.e., total area/survey area, which means that every galaxy is counted multiple times, depending on the survey coverage). We verified that imposing a minimum number of galaxies per survey and bin provided similar results, given that a small number of objects in a small survey area implies a high weight in the calculations. This occurs especially at high-$z$ and high-$M_\star$, where the number counts drop in all surveys. 

In Table~\ref{tab:bins} we show the number of objects per $M_\star$ interval and survey for each $z$ range, as well as the contribution of each survey to the bin after weighting by the survey area. This method allows us to combine datasets with different $M_\star$ completeness levels. For instance, at $1\leq z <1.5$, CANDELS 80\% completeness limit lies at $10^{8.6}\,M_\odot$, but we are also selecting galaxies with 1 dex lower $M_\star$. In this case, these galaxies contribute 3\% to the sample within the $10^{7.6 - 8.0}\,M_\odot$ interval, whereas JADES + CEERS + CANUCS + UNCOVER sum up 82\% of the galaxies that are being considered.
The distribution of the properties of our final sample is displayed as solid gray histograms in Figs. \ref{fig:mass}, \ref{fig:Av}, and \ref{fig:sfr}, and will be used to explore the MS in Sect.~\ref{sec:results}.

\subsection{Completeness}
\label{sec:completeness}

\subsubsection{Estimating completeness limits from observables}

To derive our $M_\star$ completeness, we followed two complementary approaches.  
The first is based on the evolution of the number of post-starburst galaxies with $M_\star$ and $z$, which assumes that these galaxies are the ones prone to being missed when completeness falls toward lower $M_\star$.
We created mock SEDs drawn from our \texttt{Dense Basis} atlas, choosing a post-starburst-like SFH, and scaling them to cover the whole $M_\star$ and $z$ ranges. Then, we looked for galaxies in our catalogs that resemble these mock SEDs by computing cosine similarities (i.e., comparing their SED shapes) and average ratios (i.e., comparing their continuum levels). In other words, for each of these mock SEDs (i.e., per $z$ and 0.2 dex $M_\star$ bin), we counted the number of counterparts in our catalogs. This method provides us with a selection function whose peak can be used to infer a completeness estimate at each $z$. However, as we move to higher $z$ (i.e., $z>5$), the S/N drops and challenges this method. For that reason, completeness at $z>5$ was extrapolated from a fit to the $z<5$ completeness curve, assuming that completeness decreases monotonically with $z$ following a logarithmic trend. Further details about this method will be provided in Mérida et al. (in prep), which will also discuss the impact of gravitational lensing in this computation. Considering the limits achieved by the JWST surveys included in this work, the average 80\% $M_\star$ completeness is reached at $10^{7.6}\,M_\odot$ at $z = 1$ ($10^{8.8}\,M_\odot$ at $z = 9$).

The second is a more traditional approach that relates our galaxy distribution with different stellar mass functions (SMFs; i.e., \citealt{Santini2012}, \citealt{Ilbert2013}, \citealt{Tomczak2014A}, \citealt{Grazian2015}, \citealt{Weibel2024}). The ratio
between the SMF and our $M_\star$ distribution allows us to define the completeness (see also M+23). At $z>5$, this procedure should also be used with caution until we achieve a better understanding of the SMF at this and higher $z$. The method provides similar estimates to our first approach within 0.1~dex. Our dataset allows us to push traditional mass-completeness limits (e.g., CANDELS; B+19) $\sim 1$~dex toward lower $M_\star$. We can probe $10^{8.3}\,M_\odot$ galaxies up to $z\sim4$. This implies that we cannot reach the low-$M_\star$ regime (i.e., $\lesssim10^{8}\,M_\odot$) at higher $z$. Nevertheless, we are able to push the MS much deeper than pre-JWST MS studies at these higher $z$. 

For the SFR completeness, we selected those bands centered at the rest-frame ultraviolet (UV) for each galaxy. The limiting magnitudes for the bands probing the spectral range around 280~nm rest-frame were translated into a UV luminosity and then into the SFR using the \citet{Kennicutt1998} relation (M+23). Only galaxies above the $M_\star$ limits were considered, since our goal is to obtain a sample that is both $M_\star$ and SFR complete. We considered two cases: $A_v=0$ mag and $A_v$ equal to the median value within the corresponding $z$ interval.

\begin{table}[htp]
\setlength{\tabcolsep}{1.3pt}
    \centering
    \small
    \caption{Stellar mass and SFR 80\% completeness limits.}
    \begin{tabular}{c|c|c|c}
    $z$ & log $M_\star$ [$M_\odot$]& log SFR$_{Av = 0}$ [$M_\odot$ yr$^{-1}$]& log SFR$_{\overline{Av}}$ [$M_\odot$ yr$^{-1}$]\\
    
    \hline\hline
    $1\leq z<1.5$ & 7.6 & $-1.37$& $-1.17$\\ \hline
    $1.5\leq z<2$ & 7.8 & $-1.09$& $-0.86$\\ \hline
    $2\leq z<2.5$ & 8.0 & $-0.73$& $-0.51$\\ \hline
    $2.5\leq z<3$ & 8.1 & $-0.38$& $-0.20$\\ \hline
    $3\leq z<4$ & 8.3 & $-0.02$& $0.13$\\ \hline
    $4\leq z<5$ & 8.4 & $0.31$& $0.43$\\ \hline
    $5\leq z<7$ & 8.6 & $0.92$& $1.03$\\ \hline
    $7\leq z\leq 9$ & 8.8 & $1.26$& $1.39$\\ \hline
    \end{tabular}
    \label{tab:completeness}
\tablefoot{We include the SFR values computed based on $Av=0$ (third column) and on the median $Av$ of the $z$ bin (fourth column).}
\end{table}

\subsubsection{The underlying completeness function}

Figures \ref{fig:mass} and \ref{fig:sfr} show our $M_\star$ and SFR completeness limits as vertical lines, whose values can also be consulted in Table~\ref{tab:completeness}. Both limits increase with increasing $z$. At $1<z<3$, the peak of the gray histograms roughly coincides with the SFR and $M_\star$ completeness limits. At higher $z$, the limits deviate from the peak toward higher values of both SFR and $M_\star$. The latter effect could be partially induced by the evolution in the shape of the completeness function with $z$.

\begin{figure*}[htp]
    \centering
    \includegraphics[width=\linewidth]
    {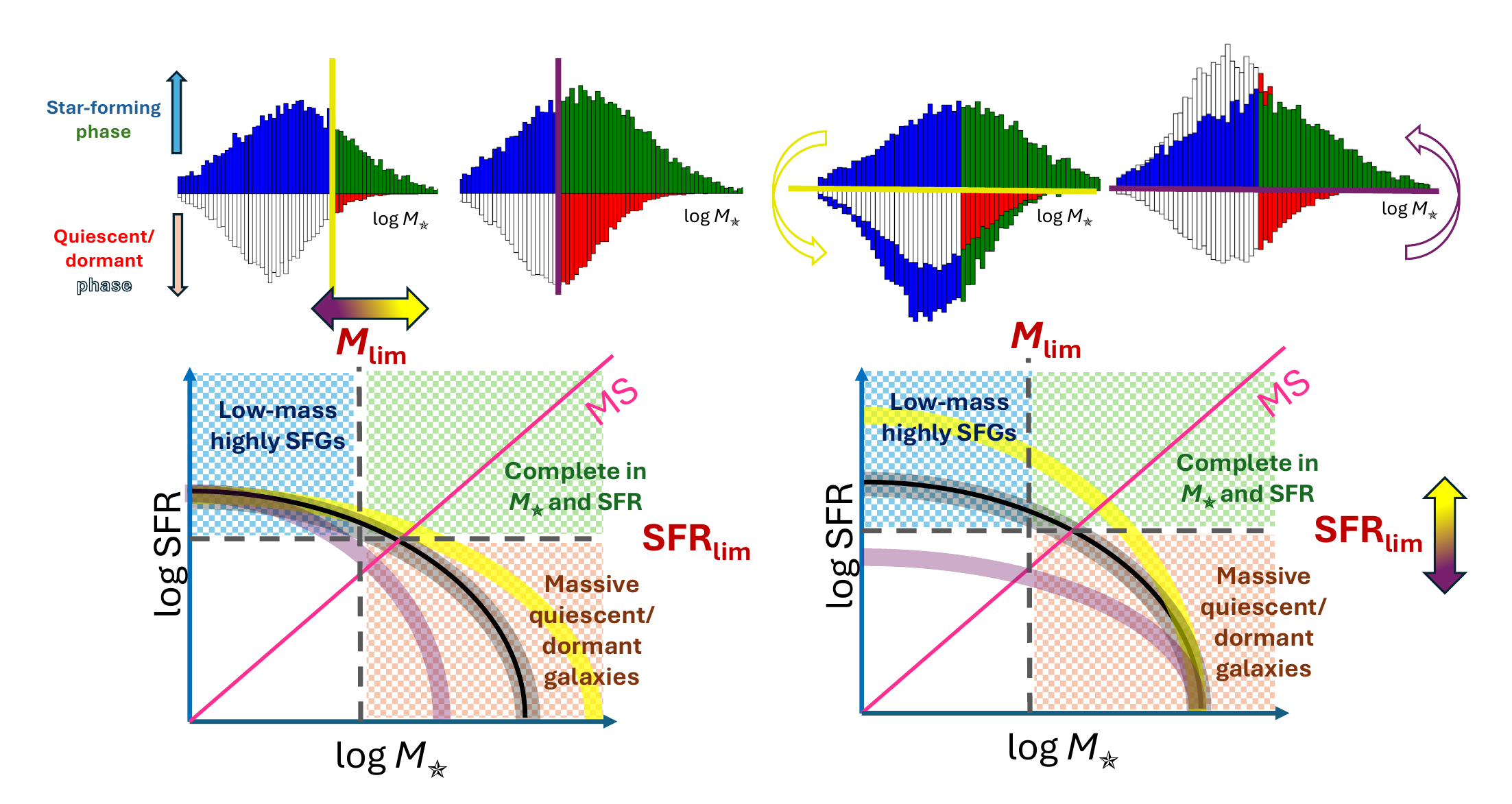}
    \caption{Cartoon depicting the variations induced in the $M_\star$ (left) and SFR (right) completeness limits by changes in the shape of the completeness function (black line). We plot some hypothetical SFR and $M_\star$ limits as dashed lines. The hatched green region encloses the SFR and $M_\star$ values above both completeness limits. The hatched blue region highlights the locus of galaxies that would be detected thanks to their high SFR, but that lie below the $M_\star$ limit. The hatched orange region contains galaxies above the $M_\star$ limit that would be hard to detect because of their low SFR. The white region is occupied by galaxies that are below the $M_\star$ and SFR limits. In yellow (pink), we show how the curve should change in order to increase (decrease) both the $M_\star$ and SFR limits independently. Histograms above each panel exemplify how changes in $M_\star$ or SFR completeness would affect our final sample. Bins are color-coded according to the regions they would occupy in the log SFR$-$log $M_\star$ plane. We display two mirrored histograms, showing galaxies that lie above (blue and green) and below (white and red) the SFR limit. The quiescent or dormant histogram is centered at a lower $M_\star$, as bursty star formation rules at lower $M_\star$ (i.e., quiescence is not permanent and these galaxies should be included in an MS analysis). At higher $M_\star$ galaxies that fall below the MS are more prone to remain quiescent, and thus would be discarded through $UVJ$ and sSFR cuts. Vertical and horizontal lines are color-coded following the bottom panels. The last two histograms exemplify how objects would move from the top to the mirrored histograms, and vice versa, as we change the SFR limit.}
    \label{fig:completeness}
\end{figure*}

These vertical lines were computed at the 80\% level. Different levels will probe different regions of the underlying completeness function, which depends on both $M_\star$ and SFR, and evolves with $z$. Figure~\ref{fig:completeness} shows a cartoon where we represent the $M_\star$ and SFR completeness limits as dashed lines, computed at a random x\% level for illustrative purposes. The underlying function from which these limits are extracted is depicted as a solid black line. The exact parametrization of this function will be explored in detail in Mérida et al. (in prep).

The composite of the completeness function and the completeness limits yields four different galaxy regimes. Objects can be sufficiently massive and star-forming to exceed the $M_\star$ and SFR limits (hatched green region). Any MS or burstiness analysis should only consider these galaxies. 
On the other hand, there are sources that are not massive enough, but whose high SFR will make them detectable (hatched blue region). Objects that are massive enough, but have a too low SFR (hatched red region) will not be detected. Finally, there is a regime occupied by galaxies that are neither massive nor star-forming enough to be detected (white region). These last two regimes are occupied by quiescent sources, whether permanently or temporarily. 

It is important to keep in mind that this is a simplified view of the parameter space. For example, some galaxies in the hatched blue region lie below the completeness curve (see the left panel of Fig.~\ref{fig:completeness}). Depending on the percentage used to compute the limits (the dashed horizontal line), the number of SFGs that skip our selection can increase or decrease. For illustrative purposes, we draw the completeness limits so that only galaxies in the white region fall below the line. In a realistic context, however, below the SFR completeness limits, we should expect to find a mixture of quiescent galaxies and some SFGs whose emission is too faint, and that thus remain undetected.

Figure~\ref{fig:completeness} also shows the impact of the completeness shape on our datasets. We depict the effect induced in the final galaxy distribution (histograms above), which has to be complete in both SFR and $M_\star$ (green). The amount of low-mass, highly star-forming galaxies (blue) below our $M_\star$ completeness and detectable SFR will be smaller if the slope of the completeness function at lower $M_\star$ becomes steeper (yellow curve in the second panel). Understanding the completeness function in terms of both $M_\star$ and SFR is thus key to avoiding biases induced by this galaxy population (see also \citealt{Simmonds2025}). 

On the other hand, at high redshifts, galaxies become increasingly faint, and their rest-frame UV emission is difficult to measure, which affects SFR completeness estimates. However, SFRs from these galaxies can be recovered through SED fitting, taking advantage of optical constraints. 

\section{The $M_\star$ versus SFR relation at low masses}
\label{sec:results}

\begin{figure*}[htp]
    \centering
\includegraphics[width=\textwidth, height=17cm]{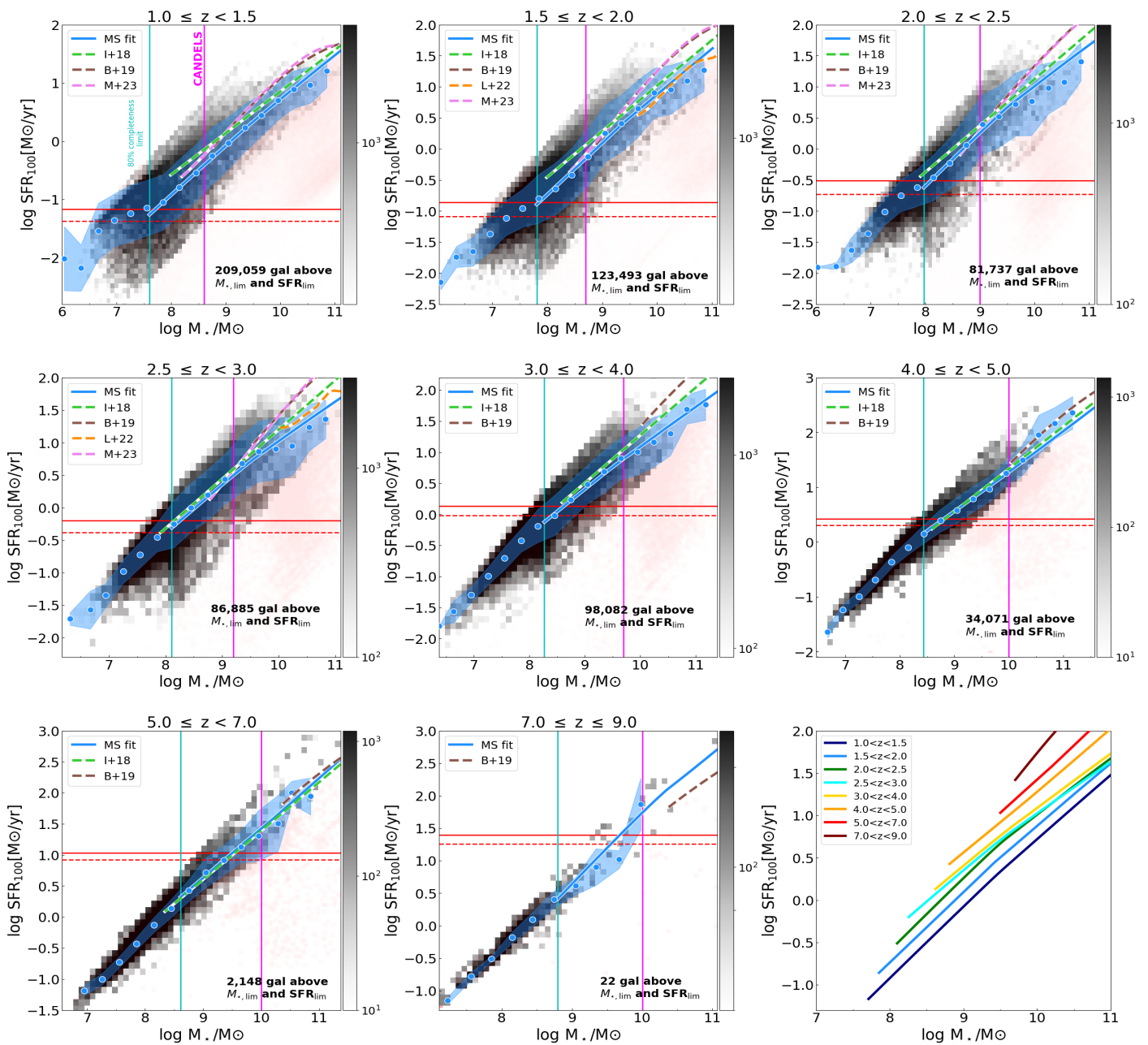}
    \caption{Star formation rates versus stellar masses for our sample, divided into different redshift intervals covering ${1\leq z\leq 9}$. Galaxies are represented in gray in a 2D histogram using a symmetric logarithmic scale. In light red, we include a scatter plot (i.e., individual points) showing the galaxies that did not pass our sSFR screening. The vertical cyan lines denote the 80\% mass completeness limits of our dataset. Previous limits from CANDELS are represented as vertical fuchsia lines (B+19, not corrected for outshining). The horizontal lines show our SFR limits, computed without attenuation (dashed red line) and taking the median $A_v$ of the sample in the corresponding $z$ interval (solid red line). The LOWESS fit to the MS is shown as a thick blue line; running medians are displayed as blue circles, and the standard deviation is represented by the shaded blue region. We include four fits from the literature as dashed lines: green for \citealt{Iyer2018}, brown for B+19, orange for \citealt{Leja2022}, and pink for M+23. These fits are plotted down to their $M_\star$ completeness limits. B+19 and M+23 fits were corrected for outshining using the \cite{Sorba2018} prescription. The last panel shows all our fits together, color-coded by redshift. They are displayed down to the corresponding $M_\star$ and SFR limits of the bin (see Sect.~\ref{sec:completeness}).}
    
    \label{fig:MS}
\end{figure*}

Figure~\ref{fig:MS} shows $M_\star$ versus SFR of our galaxy sample as grayscale 2D histograms in the redshift intervals defined previously in Sect.~\ref{sec:properties}. These two quantities exhibit a tight correlation over the examined redshift range ($1 \leq z \leq 9$).
We fit the data in each panel using the locally weighted scatterplot smoothing (LOWESS) method from the \texttt{Python}'s \texttt{statsmodels} library \citep{seabold2010statsmodels}, which is a nonparametric method. It is based on robust locally weighted regression, not assuming any global functional form \citep{Cleveland1979}. In this way, possible inflection points in the MS will not be artifacts due to the election of a concrete parametrization. 

We used this first LOWESS fit to carry out a sigma-clipping of the data ($\sigma=3$) and refit the clipped distribution with LOWESS. In both iterations, we set \texttt{frac} = 1 (i.e., using all the data). As a result of the sigma-clipping, we removed $< 1$\% of galaxies lying above and below the MS. Although this is a small number, we used the entire unclipped distribution to derive the MS parameters further in this section. 

The fit is shown in Fig.~\ref{fig:MS} as a solid blue line. On the other hand, we also binned the whole log $M_\star$ range per $z$ interval in 0.1~dex bins, computing the running median and scatter, also included in Fig.~\ref{fig:MS} (blue points and shaded blue region). We show the values of the running medians, the scatter, and the LOWESS fit in Table \ref{tab:running_medians}.

Looking at Fig.~\ref{fig:MS}, we see that at $z>5$ the number of objects above our $M_\star$ and SFR completeness limits drops. This is a consequence of imposing completeness in both SFR and $M_\star$ at the same time, in the manner detailed in Sect.~\ref{sec:completeness}. More UV-continuum faint galaxies would be required to push down completeness at $5<z<9$ (i.e., pushing down SFR completeness for this $M_\star$ limit). Additionally, wider JWST surveys are key to populating the $>10^{9}\,M_\odot$ regime at $z>7$ (i.e., pushing down $M_\star$ limits so that we reach the bulk of JWST sources; see Fig.~\ref{fig:ms_drawing}). Our findings at $z>5$ should thus be interpreted with caution until we can fill these gaps.

The normalization of the fits increases with $z$, as has been reported by previous works (e.g., \citealt{Pearson2018}, \citealt{Popesso2023}). The transition between what we call the observed low-mass regime (i.e., log ${M_{\star,\,\mathrm{lim}} < \mathrm{log}\,M_\star<\mathrm{log}\,M_{\star,\,\mathrm{CANDELS}}}$, where the CANDELS limits are those reported in B+19) and the high-mass regime (i.e., ${\mathrm{log}\,M_\star>\mathrm{log}\,M_{\star,\,\mathrm{CANDELS}}}$) is smooth. The limits of the regimes are defined to account for the redshift evolution of mass completeness.

We do not see evidence of a high-$M_\star$ turnover point, but detect a bending of these fits at lower $M_\star$. We also see that the ${2<z<2.5}$ and ${2.5<z<3}$ fits intersect at high $M_\star$. Several factors, including the MS parametrization and binning, sSFR cuts, and the treatment of uncertainties, can produce such effects. Different parameterizations and binning can reflect statistical changes in the trends not physically motivated by the data (i.e., nonphysical inflection points, see Sect.~\ref{sec:intro}). The sSFR cuts at the high-$M_\star$ end are more uncertain, as both bulge growth and AGN feedback play significant roles. We are not considering the contribution of AGN in our fits, which may induce asymmetries in the SFR uncertainties. 

We included fits from previous works; namely, \citealt{Iyer2018} (I+18), B+19, \citealt{Leja2022} (L+22), and M+23. B+19 and M+23 were corrected for outshining using the \citealt{Sorba2018} correction; still, an offset from our relation is expected, given that this is only a general correction. I+18 is based on a previous version of \texttt{Dense Basis}, which could be the underlying cause of the slight offset. L+22 is based on \texttt{Prospector}, using a nonparametric SFH, and thus more in line with our method. Overall, taking the previous factors into account, we find a good agreement between the different fits and our estimations.

To study the variations of the MS parameters with $z$ and $M_\star$ in more detail, we relied on three different methods. We used \texttt{lmfit} \citep{Newville2014} and \texttt{emcee} \citep{Foreman-Mackey2013} to run a Markov chain Monte Carlo (MCMC) sampling of the posterior distribution. For the former \texttt{Python} method, we assumed that the uncertainties in log $M_\star$ and log SFR are symmetrical. \texttt{emcee} was instead run including a split normal distribution to account for possible asymmetries in both parameters. Both methods provide compatible results when assuming symmetric uncertainties. Lastly, we computed the MS parameters using the \texttt{curve\_fit} method from \texttt{scipy.optimize} \citep{Virtanen2020}. Each of these methods, which assume a linear parametrization of the MS in log$-$log space, provides us with values for the normalization ($\alpha$), slope ($\beta$), and scatter ($\epsilon$) of the MS. We used \texttt{curve\_fit} to obtain an estimate of the observed or statistical MS parameters, which we then compared with the intrinsic values derived following an MCMC approach with \texttt{lmfit} and \texttt{emcee}.

\begin{figure*}[htp]
    \centering
    \includegraphics[width=.9\linewidth]{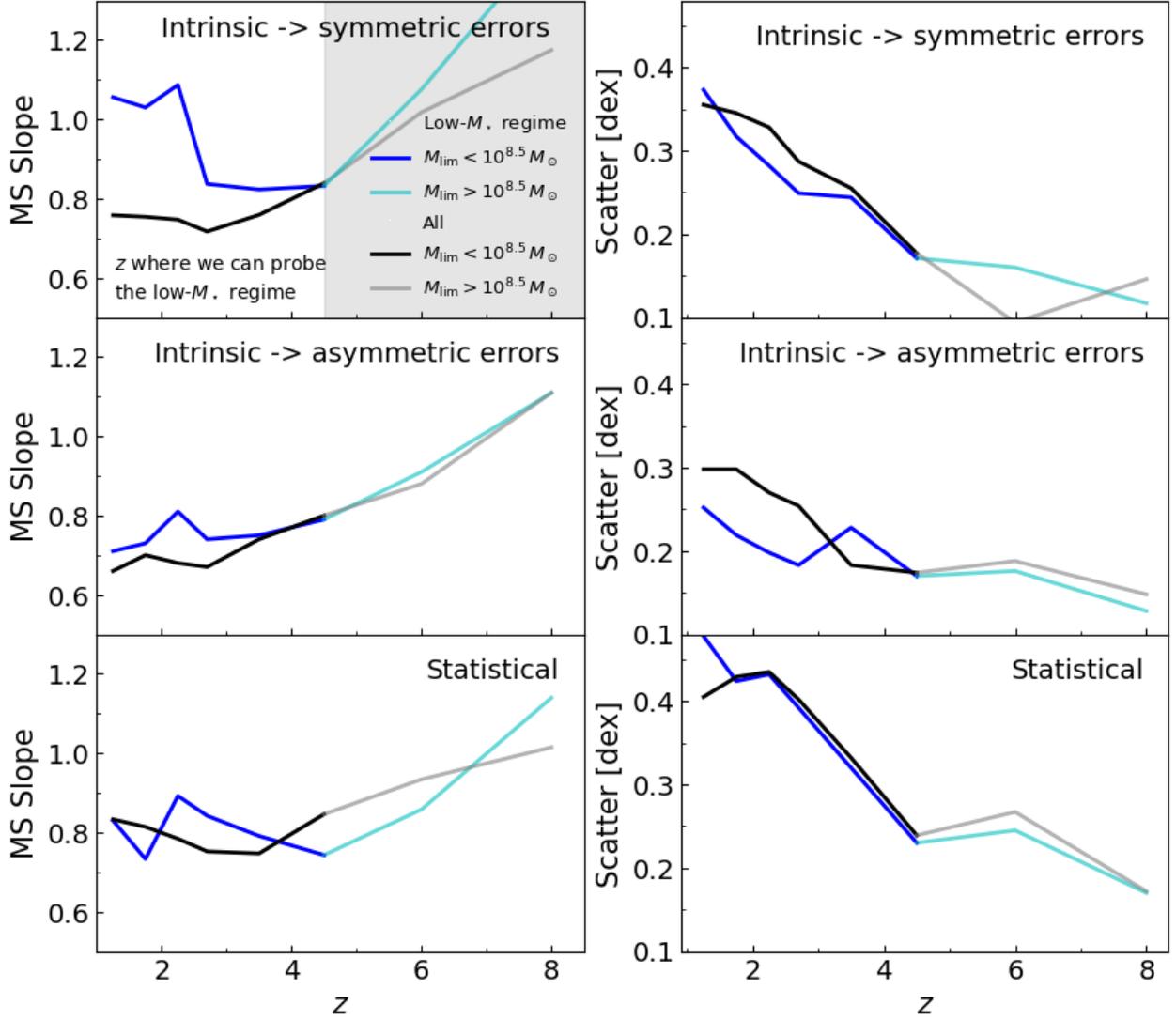}
    \caption{Main sequence slope (first column) and scatter (second column) derived using the different fitting methods; namely, \texttt{lmfit}, assuming symmetric errors (top), \texttt{emcee}, considering asymmetries (middle), and \texttt{scipy.optimize} (bottom), fitting the data to a straight line. In each subplot, blue colors show the evolution of the parameters for the low-$M_\star$ regime (i.e., below the CANDELS $M_\star$ limits), whereas black represents our findings for the overall $M_\star$ regime. Colors are more intense up to $z\sim5$, where we have the means to explore $\lesssim10^{8.5}\,M_\star$ objects with an SFR and mass-complete sample. At higher $z$, the low-$M_\star$ sample is not actually probing the physical low-$M_\star$ regime; the sample is dominated by the high-$M_\star$ end. The values here represented can be consulted in Table~\ref{tab:slopes}.}
    \label{fig:MS_parameters}
\end{figure*}

Figure \ref{fig:MS_parameters} shows our results, computed using the whole $M_\star$ regime and only considering the low-$M_\star$ regime (i.e., below the CANDELS completeness limits). The values shown in this plot can be found in Table~\ref{tab:slopes}. Focusing first on the estimates computed for the whole $M_\star$ regime (in black in Fig. \ref{fig:MS_parameters}), we see that the slopes derived considering asymmetric errors correspond to the lowest values. The difference between the \texttt{emcee} and \texttt{lmfit} methods is always $<0.10$ at $z<5$, which points to the uncertainties being overall symmetrical. The values derived using \texttt{curve\_fit} are always larger and closer to those obtained with \texttt{lmfit}. The MS slope is $\sim0.7 - 0.8$ at least up to $z\sim5$. The scatter found by \texttt{emcee} is always lower than that reported by \texttt{lmfit} (likely because the posterior is non-Gaussian), but still compatible with the canonical values reported in \citealt{Speagle2014} (i.e., $0.2 - 0.3$~dex) up to $z\sim5$. The pure statistical scatter values are larger by $\sim0.1$ dex. In Appendix~\ref{app:MS_tests} we explore the possible source for the decreasing overall scatter with $z$, finding that it is likely not physical but rather driven by the low-S/N SEDs at increasing $z$. When the S/N decreases, the posterior has a larger contribution from the prior.
In Appendix~\ref{app:MS_tests} we also include the case in which only data from JWST surveys are considered, comparing with fits from the literature based on JWST, and checking the effect on the MS parameters.       

In the low-$M_\star$ regime, the difference between the \texttt{lmfit} and \texttt{emcee} slopes is more significant. When considering asymmetric errors, we get smaller values, more in line with what was found for the whole $M_\star$ regime. A steepening of the MS could be inferred, especially looking at the \texttt{lmfit} method, at least up to $z\sim5$, the limit at which we can really probe the physical low-$M_\star$ regime. We estimated the level of asymmetry in the uncertainties' distribution (i.e., how much they deviate from a Gaussian behavior) as ${\Delta \mathrm{x}_{\mathrm{top}} - \Delta \mathrm{x}_{\mathrm{bottom}} / \Delta \mathrm{x}_{\mathrm{top}} + \Delta \mathrm{x}_{\mathrm{bottom}}}$, with “top” being the difference between the 84th and 50th percentiles and “bottom” being the difference between the 50th and 16th percentiles. It is larger in the low-$M_\star$ regime, especially in log SFR, reaching a value of $\sim - 0.30$. The latter implies that \texttt{Dense Basis} posteriors are skewed toward lower SFR (and also lower $M_\star$) values when fitting these low-mass sources. When taking asymmetries into account, the steepening thus becomes more subtle, highlighting the need to consider the nature of the uncertainties' distributions when computing intrinsic parameters. In Sect.~\ref{sec:discussion}, we discuss the physical implications of this steepening of the MS slope at low $M_\star$.

As we move to higher $z$, the differences between \texttt{lmfit} and \texttt{emcee} are reduced, as we stop probing the actual low-$M_\star$ regime and asymmetries diminish. On the other hand, the values and evolution of the scatter are similar to what we found for the whole $M_\star$ regime. \texttt{emcee} provides lower estimates for the scatter at $z<3$ at low masses, although these values lie within $0.2 - 0.3$ dex. 

\begin{table*}[]
\caption{Main sequence parameters.}
\setlength{\tabcolsep}{10pt}
    \centering
    \small
    \begin{tabular}{c|c|c|c|c|c|c|c|c|c|c}
    \hline
    \multicolumn{11}{c}{log SFR = log $M_\star\times\beta$ + $\alpha$}\\ \hline
      $z$ bin   &  & \multicolumn{3}{c|}{Instrinsic - symm} & \multicolumn{3}{c|}{Intrinsic - asymm} & \multicolumn{3}{c}{Statistical} \\ \hline
      & &$\alpha$&$\beta$&$\epsilon$&$\alpha$&$\beta$&$\epsilon$&$\alpha$&$\beta$ &$\epsilon$\\
      \hline
      $1\leq z < 1.5$ & All galaxies & $-6.81$&0.76 & 0.36& $-5.84$&0.66 & 0.30 & $-7.57$&0.83 & 0.41 \\
         & Low-$M_\star$ & $-9.22$&1.06 & 0.37 &$-6.22$& 0.71 & 0.25 & $-7.56$&0.83 & 0.48 \\
         \hline
      $1.5\leq z < 2$ & All galaxies & $-6.63$&0.75 & 0.35& $-6.05$&0.70&0.30&$-7.27$&0.81&0.43\\
         & Low-$M_\star$ &$-8.89$& 1.03 & 0.32 & $-6.23$&0.73&0.22&$-6.62$&0.73&0.42 \\
         \hline
    $2\leq z < 2.5$ & All galaxies & $-6.44$&0.75 & 0.33& $-5.68$&0.68 & 0.27 & $-6.87$&0.78 & 0.44 \\
         & Low-$M_\star$ & $-9.28$&1.09 & 0.28 & $-6.79$&0.81 & 0.20 & $-7.77$&0.89 & 0.43 \\
         \hline
    $2.5\leq z < 3$ & All galaxies & $-6.08$&0.72 & 0.29& $-5.58$&0.67 & 0.25 & $-6.47$&0.75 & 0.40 \\
         & Low-$M_\star$ & $-7.09$&0.84 & 0.25 & $-6.18$&0.74 & 0.18 & $-6.47$&0.84 & 0.39 \\
         \hline
    $3\leq z < 4$ & All galaxies & $-6.36$&0.76 & 0.26& $-6.18$&0.74 & 0.18 & $-6.33$&0.75 & 0.33 \\
         & Low-$M_\star$ & $-6.92$&0.82 & 0.24 & $-6.27$&0.75 & 0.23 & $-6.71$&0.79 & 0.32 \\
         \hline
    $4\leq z < 5$ & All galaxies & $-6.92$&0.84 & 0.18& $-6.54$&0.80 & 0.17 & $-7.03$&0.85 & 0.24 \\
         & Low-$M_\star$ & $-6.86$&0.83 & 0.17 & $-6.52$&0.79 & 0.17 & $-6.12$&0.74 & 0.23 \\
         \hline
        $5\leq z < 7$ & All galaxies & $-8.39$&1.02 & 0.09& $-7.21$&0.88 & 0.19 & $-7.75$&0.93 & 0.27 \\
         & Low-$M_\star$ & $-8.98$&1.08 & 0.16 & $-7.51$&0.91 & 0.18 & $-7.08$&0.86 & 0.25 \\
         \hline
    $7\leq z \leq 9$ & All galaxies & $-9.87$&1.18 & 0.15& $-9.38$&1.11 & 0.15 &$-8.45$& 1.02 & 0.17 \\
         & Low-$M_\star$ & $-12.45$&1.46 & 0.12 & $-9.26$&1.10 & 0.14 & $-9.62$&1.14 & 0.17 \\
         \hline
         
    \end{tabular}
    \tablefoot{Fits to SFR$-M_\star$ using the methods described in Sect.~\ref{sec:results}. Each column shows the values of the y intercept ($\alpha$), slopes ($\beta$), and scatter ($\epsilon$). For each redshift bin, we distinguish between the global and low-$M_\star$ galaxy regimes, as defined in Sect.~\ref{sec:results}. The typical uncertainties found for $\alpha$ are $\sim 0.1$; $\sim0.01$ for $\beta$ and $\epsilon$.}
    \label{tab:slopes}
\end{table*} 

\section{Discussion}
\label{sec:discussion}

As a result of our MCMC analysis, we can neither confirm nor dismiss a possible change in the MS slope at lower $M_\star$.
However, if this hint of change in the MS slope is real, the demarcation between the two different regimes, reflected by the variation in curvature of the LOWESS fits, would be located at $\sim10^{9.5}\,M_\odot$ at ${1<z<5}$. Thus, this low-$M_\star$ turnover point would remind us of the turnover observed at the high-$M_\star$ end. 

This same transition was identified at $10^8\,M_\odot$ using the FIRE-2 zoom-in simulations in \citet{Ma2018} and observationally at $10^{8.6}\,M_\odot$ at $6 < z < 7$ in \citet{Ciesla2024}. While the exact position of the turnover could vary overall, we see that its value is independent of redshift, as in M+23, where this threshold was found at $10^{8.8}\,M_\odot$ at ${1 < z < 3}$ with no $z$ dependence. The lack of $z$ dependence could point to an event in a galaxy's lifetime that would only depend on its ability to retain gas. 

This moment could match the assembly of stable disks. 
We now know that dynamically cold disks could be present even at very high-$z$ (e.g., $z=7.3$, \citealt{Rowland2024}).
\citet{Dekel2020} affirmed that gas disks tend to survive only in haloes above a threshold mass of $\sim2 \times 10^{11}\,M_\odot$, which would correspond to $M_\star \sim 10^9 \, M_\odot$, with only a weak redshift dependence. According to this work, the threshold arises from the halo merger rate when accounting for the mass dependence of the ratio of galactic baryons and halo mass. The turnover at $M_\star\sim 10^{9}\, M_\odot$ corresponds to a halo mass of $\sim10^{11}\,M_\odot$, with a weak evolution with $z$ \citep{Behroozi2019}.

Supernova feedback is key in disrupting disks below the critical mass by driving the stellar-to-halo mass ratio that affects the merger rate, stirring up turbulence and suppressing high-angular-momentum gas supply, and confining major compactions to the critical mass (however, see \citealt{Romeo2018}, who argued that nearby star-forming spirals self-regulate to a quasi-universal disk stability level, showing how disk instabilities hardly correlate with $M_\star$).

\cite{Dekel2020} point out that if the $M_\star$ to baryonic mass ratio evolved with redshift, this could induce a $z$ dependence on the critical mass mentioned above. A decrease of this quantity with $z$ would imply a lower fraction of disks at higher redshifts at a fixed mass.

Moreover, \citet{Simons2015} found a similar turnover mass in the context of the local Tully-Fisher relation \citep{Tully1977}, which links $M_\star$ with the rotation velocity of galaxies. They used a morphologically blind selection of local, field emission line galaxies, finding a transition $M_\star$ in this relation at $\sim10^{9.5}\,M_\odot$. Above this critical mass, nearly all galaxies are rotation-dominated and are, on average, more morphologically disk-like. Below, the relation shows significant scatter to low rotation velocity, and galaxies can either be rotation-dominated disks or asymmetric or compact galaxies that scatter off. They refer to this transition mass as the “mass of disk formation”. \citet{Benavides2025}, using the FIREbox simulation \citep{Feldmann2023}, found transition masses compatible with these values, seeing that galaxies with $M_\star<10^9\,M_\odot$ are rarely disk-dominated.
Additionally, \citet{Ganapathy2025} characterized the asymmetry of 271 disk galaxies within CEERS at ${1 < z < 4}$, finding that disk galaxies with lower masses (reaching down to $M_\star = 10^{8.2}\,M_\odot$) tend to be more asymmetric. They measured the relation between disk galaxy asymmetry and $z$, finding no conclusive relationship between them. 

A detailed study of the evolution of galaxy morphology with mass and redshift using JWST data could shed light on this transition, its duration, and dependencies, since the environment may also play an important role. However, it is still difficult to delve deeper into the morphology of $<10^9\,M_\odot$ galaxies. Given that sources extend over smaller angular sizes, there is an increased effect from the PSF, and sources are dimmer due to
cosmological dimming \citep{Ferreira2023}. \citet{Huertas-Company2024} classify the majority of low-mass galaxies in CEERS as irregulars: less bulge-dominated systems at all redshifts. As is pointed out in that same study, fluctuations in the surface brightness caused by noise can be easily interpreted as an irregular light distribution. Based also on CEERS galaxies, in this case at ${1.5<z<6}$, \citet{Ferreira2023} concluded that low-mass disks ($<10^9~M_\odot$) are more likely to be peculiar galaxies with unresolved structure at all $z$. Deeper observations will better detect potential diffuse components around galaxies, moving some of these irregulars to the disk class.

Above this low-mass turnover mass, or “mass of
disk formation”, galaxies would be fully consistent with the self-regulated evolution model, becoming massive enough to counteract feedback, and thus preventing further stochastic events. Below this threshold, galaxies are on their way to acquiring enough mass to become stable against feedback (which is powerful enough to expel the gas from the galaxies, given their low gravitational potential), developing a stable disk. The lower the $M_\star$, the more stochastic and bursty the star formation will be. The steepening in the MS slope would suggest that the SFE or the gas content is lower as we move toward lower $M_\star$. In other words, we should expect to find more galaxies in a latent phase rather than in a burst phase at progressively lower $M_\star$. The latter was probed via simulations in \citet{Gelli2023}, which showed that the duty cycle gets increasingly smaller toward lower $M_\star$. In a continuation work, these findings were also compared with JWST observations, reporting consistent results \citep{Gelli2025}. At $M_\star<10^8\,M_\odot$, temporarily quiescent galaxies would represent the dominant population. Galaxies with $10^{7.5}\,M_\odot$ would show a duty cycle of $\sim0.6$, while those with $10^{9.0}\,M_\odot$ would show a duty cycle of 0.99, undergoing no more latent phases.

On the other hand, \citet{Hunt2020} found that dwarf star-forming local galaxies (that they define as those with ${M_\star<3\times10^9\,M_\odot}$) tend to be overwhelmed by HI accretion, so that star formation is not able to keep up with the gas supply. In galaxies with intermediate masses, which they define as ${3\times 10^9\,M_\odot < M_\star < 3\times10^{10}\,M_\odot}$ sources, star formation proceeds apace with gas availability, and HI and H$_2$ are both proportional to SFR. These findings support the lower SFE in low-mass galaxies, and thus their rising number in latent phases as we go to a progressively lower $M_\star$ regime.

\citet{Davies2025} studied the movements of ${z<1}$ galaxies in the sSFR$-M_\star$ plane using the Deep Extragalactic VIsible Legacy Survey (DEVILS). According to this work, low-mass sources display a mix of rapidly increasing SFHs at high sSFRs (i.e., bursty galaxies), and constant SFHs (i.e., they form stars at a constant rate, but this rate is dependent on a quantity other than just $M_\star$) that remained similar for the entirety of the life of the galaxy. They point out that a high SFR dispersion at these masses could be produced by a combination of stochastic star formation processes and galaxies with a large variation in normalization of constant SFHs. In the same line, \citet{Simmonds2025}, exploring the MS using JADES data at ${3<z<9}$, pointed out that shorter-term fluctuations dominate the MS scatter, although long-term variations in star formation activity are also present.
\citet{Mintz2025} studied JWST/PRISM spectra of 41 low-mass ($<10^9\,M_\odot$) galaxies at ${1<z<3}$, many of them showing Balmer breaks. In contrast, they found that star-forming bursts occur on long timescales, $\geq100$ Myr, after which galaxies fall well below the MS. They also claim that, unlike what happens at low-$z$, high-$z$ star-forming and dormant galaxies should be considered a unified population (i.e., the concepts of red and blue clouds, or hatched red and green regions in Fig.~\ref{fig:completeness}, do not apply for low-$M_\star$ and/or high-$z$ galaxies). In other words, dormant states of galaxies are not permanent. Rather, galaxies can overcome them and enter bursty phases. \citet{Sun2023} showed that the observability of low-$M_\star$ objects is highly time-dependent when they are close to the survey’s limiting flux. As a consequence of the SFR variability, the same galaxy oscillates in and out of the observable sample. Looking for statistical samples of dormant low-$M_\star$ galaxies is key to understanding the evolution of low-$M_\star$ sources and providing better constraints on their SFHs, the MS, and other scaling relations.

\begin{figure}
    \centering
    \includegraphics[width=\linewidth]{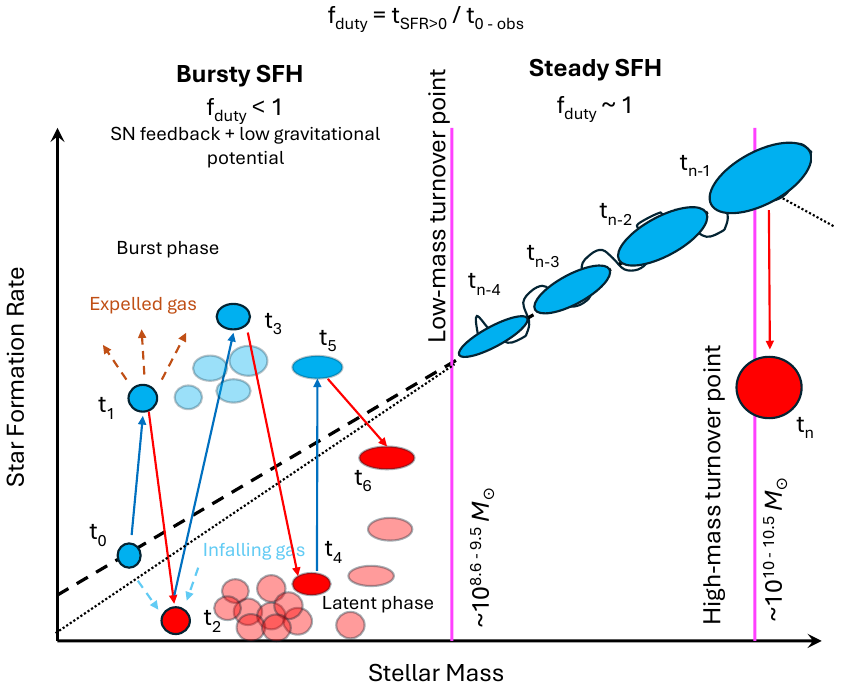}
    \caption{Cartoon depicting a possible evolutionary pathway of low-$M_\star$ galaxies along the main sequence. Galaxies in a burst or star-forming phase are shown with blue colors. Those in red correspond to latent or dormant phases or quiescence (the distance of these galaxies from the main sequence can be much larger). The vertical lines denote the high-$M_\star$ turnover point from the literature (see Sect.~\ref{sec:intro}) and the low-$M_\star$ turnover point reported in this work, which is $ z$-independent. We include an $M_\star$ range that covers the different values reported from the literature and our estimates. Our value should be regarded as the end of the transition from a busty stochastic SFH to the smooth accretion mode. Below the low-$M_\star$ turnover point, the star formation is mainly driven by supernova feedback that can expel the gas from the galaxy and push it to a latent phase. Infalling gas reignites star formation again in the form of a new burst. As the galaxy assembles mass, the disk starts to form, and the galaxy becomes more stable against feedback. The number of galaxies in a latent phase increases with decreasing $M_\star$, which can be reflected in a steepening of the main-sequence slope.}
    \label{fig:summary}
\end{figure}

To this complex picture, we should also add possible effects of AGN feedback on low-mass galaxies \citep{Martin-Navarro2018}, the presence of dusty dwarfs \citep{Bisigello2023}, and the impact of galaxy mergers (\citealt{Asada2023, Asada2024}, \citealt{Merida2025}), which are key in driving burstiness.
Finally, the fact that the scatter we report is consistent with that found at higher $M_\star$ in the literature may suggest that the stochasticity induced by the bursty SFHs undergone by low-mass galaxies could show its imprint at lower $M_\star$ than the values probed in this work, especially as we move toward higher $z$. It is more likely, though, that the scatter due to the bursts may be suppressed enough that the global SFRs average out over $\sim100$~Myr timescales, not showing up in our results (see \citealt{Simmonds2025}). \citet{Caplar2019} presented the scatter along the MS retrieved based on different SFR timescales and power spectrum densities, characterized by a high-frequency slope ($\alpha)$ and a break timescale ($\tau_{\mathrm{break}}$). Assuming bursty values for these parameters, they showed that SFR$_{10}$ provides a wide $\Delta$MS (i.e., log SFR $-$ log SFR$_{\mathrm{MS}}$), while the SFR$_{100}$ scatter remains much tighter. As the break timescales increase and $\alpha$ decreases (i.e., we get closer to a smooth SFH), the $\Delta$MS values provided by both tracers become more similar.
Figure~\ref{fig:summary} sketches a summary of our interpretation of the behavior of the MS in the low-$M_\star$ galaxy regime based on the previous information.

\section{Summary and conclusions}
\label{sec:conclusions}
  
In this work, we combined pre-JWST surveys (i.e., CANDELS, COSMOS2020, and SHARDS/CANDELS FAINT) with JWST surveys (i.e., CEERS, JADES, CANUCS, and UNCOVER) to examine the evolution of the main sequence at ${1 \leq z \leq 9}$. This comprehensive sample allowed us to explore this scaling relation across a broad range of stellar masses and SFRs, providing insights into the inflection points that reflect the transition between the different mass regimes.

The sample, made up of 755,196 sources, was refit to obtain consistent SFRs and $M_\star$ using the \texttt{Dense Basis} code and a unified configuration to avoid biases, choosing a nonparametric SFH setting. This dataset allowed us to reach down to $10^{7.6}\,M_\odot$ at $z = 1$ ($10^{8.8}\,M_\odot$ at $z = 9$) with an 80\% $M_\star$ completeness, providing the means to investigate the main sequence in the low-$M_\star$ regime up to $z\sim5$ and in general up to $z\sim9$. Below the $M_\star$ and SFR limits (which must be imposed simultaneously) and toward higher $z$, we are more susceptible to being biased toward highly star-forming galaxies in a burst phase, while unable to select the dormant or latent population of low-$M_\star$ galaxies. Conclusions regarding the main sequence's behavior or burstiness should be considered with caution until we can actually probe the complex SFR path traversed by low-$M_\star$ galaxies.

The MS slopes and scatter we obtain are compatible with canonical values found in the literature up to $z\sim5$ ($\beta\sim0.7-0.8$, $\epsilon\sim0.2-0.3$ dex). When dividing the $M_\star$ range to focus on low-$M_\star$ sources, we can detect hints of a steepening of the MS slope. This effect is subtle when considering the asymmetric nature of uncertainties in this mass regime. The steepening starts at $\sim10^{9.5}\,M_\odot$, independently of redshift, which could be reflecting a critical value above which a stable disk finally assembles in galaxies. Below that mass, galaxies would be increasingly burstier and more stochastic, and likely undergo long periods of latency below the main sequence. Extending this study toward lower $M_\star$ while achieving completeness is a complex task that requires capturing galaxies in such dormant stages.

The scatter found in the low-$M_\star$ regime is also compatible with the canonical values. In this case, the use of a long-term SFR indicator may be smoothing out rapid fluctuations suffered by low-mass galaxies, fluctuations that may be better captured by shorter-timescale SFR tracers.

\begin{acknowledgements}

This research was enabled by grant 18JWST-GTO1 from the Canadian Space Agency and Discovery Grant and Discovery Accelerator funding from the Natural Sciences and Engineering Research Council (NSERC) of Canada to MS. MB acknowledges support from the ERC Grant FIRSTLIGHT, Slovenian national research agency ARIS through grants N1-0238 and P1-0188, and the program HST-GO-16667, provided through a grant from the STScI under NASA contract NAS5-26555.\\

This research used the Canadian Advanced Network For Astronomy Research (CANFAR) platform, operated in partnership by the Canadian Astronomy Data Centre and The Digital Research Alliance of Canada with support from the National Research Council of Canada, the Canadian Space Agency, CANARIE, and the Canada Foundation for Innovation.\\

This work is partially based on observations taken by the CANDELS Multi-Cycle Treasury Program with the NASA/ESA HST, which is operated by the Association of Universities for Research in Astronomy, Inc., under NASA contract NAS5-26555.
This work is also based on observations made with the NASA/ESA/CSA \textit{James Webb} Space Telescope. The data were obtained from the Mikulski Archive for Space Telescopes at the Space Telescope Science Institute, which is operated by the Association of Universities for Research in Astronomy, Inc., under NASA contract NAS 5-03127 for JWST. These observations are associated with programs ERS-1345; GTO-1180, 1181, 1210, 1286, GO-1895, 1963; GTO-1208, GO-3362; and GO-2561. This research is also based on observations collected at the European Southern Observatory under ESO programme ID 179.A-2005 and on data products produced by CALET and the Cambridge Astronomy Survey Unit on behalf of the UltraVISTA consortium.

\end{acknowledgements}

\bibliography{aa57448-25}
\bibliographystyle{aa}

\appendix

\section{Survey descriptions}
\label{app:surveys}

In this appendix we provide more details about the surveys being combined in this work.

\subsection{CANDELS}

CANDELS is a 902-orbit HST Multi-Cycle Treasury program that combined previously obtained multi-wavelength data, including Advanced Camera for Surveys (ACS) optical observations, with new Wide Field Camera 3 (WFC3) near infrared (NIR) images. The project comprises five sky fields: the Great Observatories Origins Deep Survey fields (GOODS-N and GOODS-S, \citealt{Dickinson2003}, \citealt{Giavalisco2004}), the UKIDSS Ultradeep Survey field  (UDS, \citealt{Lawrence2007}, \citealt{Cirasuolo2007}), the Extended Groth Strip (EGS, \citealt{Davis2007}), and the Cosmological Evolution Survey field (COSMOS, \citealt{Scoville2007}).

The exploitation of these data yielded multi-wavelength photometric catalogs, spanning from the UV to the far infrared, for each of these fields (\citealt{Guo2013} for GOODS-S; \citealt{Galametz2013} for UDS; \citealt{Nayyeri2017} for COSMOS; \citealt{Stefanon2017} for EGS; and B+19 for GOODS-N). The CANDELS source selection was based on the emission in the $F160W$ NIR filter. These observations enabled the detection of over 250,000 galaxies.

\subsection{COSMOS2020}

The COSMOS2020 catalog comprises 1.7 million galaxies across the 2 deg$^2$ of the COSMOS field, $\sim$966,000 of them measured with all available broad-band data. COSMOS2020 adds deeper optical and NIR images from the Subaru Hyper Suprime-Cam and VISTA Infrared Camera surveys, as well as the definitive reprocessing of all \textit{Spitzer} data ever taken on COSMOS. It also includes all archival MegaCam COSMOS $U$ data, recombined in addition to new data taken as part of the Canada France Hawaii Telescope Large Area $U$-band Deep Survey (CLAUDS, \citealt{Sawicki2019}).
Source selection in this survey was carried out on a chi-squared $izYJHKs$ image \citep{Szalay1999}.

\subsection{SHARDS/CANDELS FAINT}

The SHARDS/CANDELS FAINT catalog (M+23) consists of 34,061 ultra-faint ($27< i < 30$ mag) galaxies at $1< z <3$ detected in the GOODS-N field. The sample was built by stacking the 25 ultra-deep medium-band images gathered by the Gran Telescopio de Canarias Survey for High-z Absorption Red and Dead Sources project (GTC/SHARDS; \citealt{Perez-Gonzalez2013}) as well as the ACS broad-band observations from CANDELS. SHARDS obtained data in 25 medium-band filters in the 500–950 nm spectral range, reaching at least magnitude 26.5 at the 5$\sigma$ level at subarcsecond seeing in each one and providing a spectral resolution of R = 50 in the optical.

\subsection{CEERS}

The Cosmic Evolution Early Release Science Survey (CEERS; ERS-1345, \citealt{Finkelstein2017}, \citealt{Finkelstein2022}, \citealt{Bagley2023}, \citealt{Finkelstein2025}) is a 77.2 hrs Director's Discretionary Early Release Science Program that consists of NIRCam and Mid-Infrared Instrument (MIRI, \citealt{Rieke2015}) imaging, Near Infrared Spectrograph (NIRSpec, \citealt{Jakobsen2022}) low (R~$\sim100$) and medium (R~$\sim1000$) resolution spectroscopy, and NIRCam slitless grism (R~$\sim1500$) spectroscopy. It targeted the HST-observed region of EGS, supported by a rich set of multiwavelength observations.
Specifically, this survey counts on NIRCam imaging in six broadband filters ($F115W$, $F150W$, $F200W$, $F277W$, $F356W$, and $F444W$) and one medium-band filter ($F410M$) over four pointings, obtained in parallel with primary MIRI observations. Source identification was based on the emission in the $F277W$ NIRCam band. The CEERS photometric catalog contains $\sim100,000$ galaxies \citep{Merlin2024}.

\subsection{JADES}

The \textit{James Webb} Space Telescope Advanced Deep Extragalactic Survey (JADES; GTO-1180, 1181, 1210, 1286, GO-1895, 1963, \citealt{Eisenstein2023}) is a program of infrared imaging and spectroscopy in the GOODS-S and GOODS-N deep fields, designed to study galaxy evolution from high $z$ to cosmic noon. It uses $\sim770$ hrs of Cycle 1 NIRCam and NIRSpec guaranteed time. In GOODS-S, in and around the \textit{Hubble} Ultra Deep Field and \textit{Chandra} Deep Field South, JADES produced a deep imaging region of $\sim$45 arcmin$^2$ (average of 130 hrs of exposure time spread over 9 NIRCam filters). This is extended at medium depth in GOODS-S and GOODS-N with NIRCam imaging of $\sim175$ arcmin$^2$ (average exposure time of 20 hrs spread over $8-10$ filters). Both fields were characterized with NIRSpec multi-object spectroscopy and coordinated MIRI parallels. Source detection was based on an inverse-variance-weighted stack of the NIRCam $F277W$, $F335M$, $F356W$, $F410M$, and $F444W$ images, which led to the identification of $\sim170,000$ galaxies. In this work, we used the GOODS-S-Deep v2.0 and GOODS-N v1.0 catalogs (\citealt{Rieke2023b}, \citealt{Hainline2024}, \citealt{Eisenstein2023b}).

\subsection{CANUCS}

The CAnadian NIRISS Unbiased Cluster Survey (CANUCS; GTO-1208, \citealt{Willott2022}) is a JWST Cycle 1 GTO program targeting 5 lensing clusters and their adjacent flanking fields in parallel (Abell 370, MACS0416, MACS0417, MACS1149, MACS1423), with a total surveyed area of $\sim100$~arcmin$^{2}$. This project combines NIRCam imaging, Near InfraRed Imager and Slitless Spectrograph (NIRISS, \citealt{Doyon2023}) slitless spectroscopy, and NIRSpec prism multi-object spectroscopy. It also includes Technicolor data (TEC; GO-3362, PI: Muzzin): a Cycle 2 follow-up GO program targeting 3 of the CANUCS clusters (Abell 370, MACS0416, MACS1149). Technicolor provides slitless spectroscopy with NIRISS in $F090W$ to the cluster fields, while adding 8 wide, medium, and narrow band filters to the flanking fields. Source detection was based on a deep $\chi_{mean}$ detection image, created by co-adding background-subtracted images for all available JWST and HST-optical filters in each field. This method enabled the identification of $\sim121,000$ galaxies \citep{Sarrouh2025}.

\subsection{UNCOVER}

The Ultradeep NIRSPec and NIRCam Observations before the Epoch of Reionization survey (UNCOVER; GO-2561, \citealt{Bezanson2024}) is a Cycle 1 JWST Treasury program that includes ultradeep imaging of $\sim45$ arcmin$^2$ on and around the Abell 2744 galaxy cluster, with follow-up on $\sim500$ galaxies with NIRSpec/PRISM. The latest data release catalog ("SUPER", \citealt{Weaver2024}) was built from a selection based on a long-wavelength $F277W$+$F356W$+$F444W$ detection image with photometry measured on the 27 available HST and JWST bands over 56 arcmin$^2$. These data are enriched with several major surveys, including MegaScience \citep{Suess2024} and GLASS \citep{Treu2022}. The UNCOVER source catalog contains $\sim70,000$ galaxies. 

\section{$M_\star$ binning and weighting}
\label{app:bins}

In Table~\ref{tab:bins}, we show the number of galaxies per survey considered within each log $M_\star$ interval and redshift range. We include the weight of each survey in the overall counts. 

\begin{table*}[htp]
\setlength{\tabcolsep}{5pt}
\caption{log $M_\star$ binning per survey and redshift interval.}
    \centering
    \fontsize{9}{8.1}\selectfont 
    \begin{tabular}{c|c|c|c|c|c|c|c|c|c|c|c|c|c|c|c}
\hline\hline
log $M_\star$ 
& \multicolumn{2}{c|}{COSMOS2020} 
& \multicolumn{2}{c|}{CANDELS} 
& \multicolumn{2}{c|}{\scriptsize{SHARDS/CANDELS}} 
& \multicolumn{2}{c|}{CEERS} 
& \multicolumn{2}{c|}{JADES} 
& \multicolumn{2}{c|}{CANUCS} 
& \multicolumn{2}{c}{UNCOVER}\\

{[$M_\odot$]} 
& \multicolumn{2}{c|}{} 
& \multicolumn{2}{c|}{} 
& \multicolumn{2}{c|}{\scriptsize{FAINT}} 
& \multicolumn{2}{c|}{} 
& \multicolumn{2}{c|}{} 
& \multicolumn{2}{c|}{} 
& \multicolumn{2}{c}{} \\
\hline\hline
\multicolumn{15}{c}{$1\leq z <1.5$} \\ \hline
          $7.6\leq \mathrm{log}\, M_\star<8$&14,290&3\%&3,951&8\%&395&6\%&1,255&22\%&1,669&17\%&1,177&21\%&702&22\%\\\hline
          $8\leq \mathrm{log}\, M_\star<8.5$&40,918&12\%&7,553&17\%&42&1\%&702&15\%&880&11\%&1,057&22\%&596&22\%\\\hline
          $8.5\leq \mathrm{log}\, M_\star<9$&37,632&21\%&5,524&24\%&-&-&158&6\%&282&7\%&410&17\%&354&26\%\\\hline
          $9\leq \mathrm{log}\, M_\star<9.5$&32,348&31\%&3,096&23\%&-&-&64&4\%&168&7\%&170&12\%&191&24\%\\\hline
          $9.5\leq \mathrm{log}\, M_\star<10$&26,046&40\%&1,749&21\%&-&-&26&3\%&78&4\%&77&9\%&117&23\%\\\hline
          $10\leq \mathrm{log}\, M_\star<11$&26,516&52\%&1,215&19\%&-&-&8&1\%&48&4\%&44&7\%&65&17\%\\\hline
          \multicolumn{15}{c}{$1.5\leq z <2$} \\ \hline
          $7.8\leq \mathrm{log}\, M_\star<8$&1,680&$<1$\%&2,047&7\%&318&8\%&816&25\%&1,361&13\%&1,392&24\%&471&14\%\\\hline
          $8\leq \mathrm{log}\, M_\star<8.5$&15,794&4\%&66,043&14\%&183&3\%&1,611&28\%&1,361&13\%&1,392&24\%&471&14\%\\\hline
          $8.5\leq \mathrm{log}\, M_\star<9$&22,468&12\%&6,087&25\%&32&1\%&427&14\%&383&8\%&498&19\%&304&21\%\\\hline
          $9\leq \mathrm{log}\, M_\star<9.5$&22,129&24\%&3,201&27\%&18&1\%&100&8\%&159&7\%&129&10\%&168&23\%\\\hline
          $9.5\leq \mathrm{log}\, M_\star<10$&14,717&30\%&1,724&23\%&-&-&68&10\%&77&6\%&52&8\%&91&24\%\\\hline
          $10\leq \mathrm{log}\, M_\star<11$&12,244&33\%&1,382&29\%&-&-&23&5\%&72&8\%&31&6\%&55&19\%\\\hline
 \multicolumn{15}{c}{$2\leq z <2.5$} \\ \hline
    $8\leq \mathrm{log}\, M_\star<8.5$&9,778&3\%&4,583&12\%&428&9\%&548&14\%&905&13\%&1,079&28\%&449&20\%\\\hline
    $8.5\leq \mathrm{log}\, M_\star<9$&21,168&17\%&3,867&22\%&30&1\%&181&10\%&280&9\%&289&16\%&263&26\%\\\hline
    $9\leq \mathrm{log}\, M_\star<9.5$&16,059&24\%&1,972&22\%&26&2\%&84&9\%&91&6\%&99&11\%&142&27\%\\\hline
    $9.5\leq \mathrm{log}\, M_\star<10$&7,568&26\%&933&25\%&-&-&26&7\%&26&4\%&139&10\%&63&28\%\\\hline
    $10\leq \mathrm{log}\, M_\star<11$&4,874&27\%&554&23\%&-&-&15&6\%&15&3\%&19&8\%&47&33\%\\\hline      
 \multicolumn{15}{c}{$2.5\leq z <3$} \\ \hline
     $8.1\leq \mathrm{log}\, M_\star<8.5$&4,950&3\%&1,136&4\%&900&27\%&71&3\%&877&19\%&672&25\%&301&20\%\\\hline
     $8.5\leq \mathrm{log}\, M_\star<9$&23,362&17\%&3,537&20\%&166&7\%&48&3\%&395&12\%&318&17\%&248&24\%\\\hline
     $9\leq \mathrm{log}\, M_\star<9.5$&24,454&30\%&2,902&27\%&33&2\%&18&2\%&136&7\%&73&6\%&164&26\%\\\hline
     $9.5\leq \mathrm{log}\, M_\star<10$&13,094&33\%&1,474&29\%&23&3\%&5&$<1$\%&54&6\%&27&5\%&71&23\%\\\hline
     $10\leq \mathrm{log}\, M_\star<11$&7,277&41\%&564&24\%&-&-&-&-&37&9\%&9&4\%&32&23\%\\\hline
 \multicolumn{15}{c}{$3\leq z <4$} \\ \hline
    $8.3\leq \mathrm{log}\, M_\star<8.5$&7,528&7\%&543&4\%&-&-&52&3\%&711&26\%&448&28\%&292&33\%\\\hline
    $8.5\leq \mathrm{log}\, M_\star<9$&30,407&22\%&3,194&18\%&-&-&42&2\%&536&16\%&336&18\%&255&24\%\\\hline
    $9\leq \mathrm{log}\, M_\star<9.5$&28,996&31\%&3,774&31\%&-&-&34&3\%&197&9\%&103&8\%&136&19\%\\\hline
    $9.5\leq \mathrm{log}\, M_\star<10$&15,469&36\%&1,758&32\%&-&-&6&1\%&89&9\%&44&7\%&52&16\%\\\hline
    $10\leq \mathrm{log}\, M_\star<11$&6,853&50\%&332&18\%&-&-&2&1\%&27&9\%&15&8\%&15&14\%\\\hline
 \multicolumn{15}{c}{$4\leq z <5$} \\ \hline
     $8.4\leq \mathrm{log}\, M_\star<9$&13,938&18\%&244&$2$\%&-&-&36&3\%&529&27\%&249&23\%&163&27\%\\\hline
     $9\leq \mathrm{log}\, M_\star<9.5$&14,195&39\%&831&$17$\%&-&-&1&$<1$\%&116&13\%&77&15\%&46&16\%\\\hline
     $9.5\leq \mathrm{log}\, M_\star<10$&6,940&39\%&646&$28$\%&-&-&2&$<1$\%&43&10\%&23&9\%&19&14\%\\\hline
     $10\leq \mathrm{log}\, M_\star<11$&1,969&53\%&119&$25$\%&-&-&1&$2$\%&7&8\%&3&6\%&2&7\%\\\hline
 \multicolumn{15}{c}{$5\leq z <7$} \\ \hline
           $8.6\leq \mathrm{log}\, M_\star<9$&2,371&7\%&40&$<1$\%&-&-&56&13\%&261&34\%&87&20\%&64&26\%\\\hline
           $9\leq \mathrm{log}\, M_\star<9.5$&2,629&15\%&163&7\%&-&-&25&10\%&56&13\%&61&25\%&40&29\%\\\hline
           $9.5\leq \mathrm{log}\, M_\star<10$&1,428&23\%&222&27\%&-&-&8&9\%&13&8\%&15&17\%&8&16\%\\\hline
            $10\leq \mathrm{log}\, M_\star<11$&475&21\%&73&22\%&-&-&11&32\%&6&11\%&3&9\%&1&6\%\\\hline
 \multicolumn{15}{c}{$7\leq z \leq 9$} \\ \hline
    $8.8\leq \mathrm{log}\, M_\star<9$&2&$<1$\%&2&$<1$\%&-&-&8&25\%&22&39\%&6&19\%&3&17\%\\\hline
    $9\leq \mathrm{log}\, M_\star<9.5$&2&$<1$\%&3&$<1$\%&-&-&8&38\%&4&11\%&5&24\%&3&26\%\\\hline
    $9.5\leq \mathrm{log}\, M_\star<10$&4&$<1$\%&5&$5$\%&-&-&8&73\%&1&5\%&-&-&1&16\%\\\hline
    $10\leq \mathrm{log}\, M_\star<11$&2&$<1$\%&1&$1$\%&-&-&4&56\%&-&-&3&42\%&-&-\\\hline
\end{tabular}
\label{tab:bins}
\tablefoot{For each survey, we include the number of objects found in each $M_\star$ interval, including the contribution of each survey to that bin after weighting by the survey area. The SHARDS/CANDELS FAINT sample only covers $1<z<3$.}
\end{table*}

\section{Main sequence fit values}
\label{app:MS_fit}

In Table~\ref{tab:running_medians}, we include the running medians, statistical scatter, and the LOWESS MS fit, presented in Sect.~\ref{sec:results}.

\begin{table*}[htp]
\setlength{\tabcolsep}{1pt}
\caption{Running medians, statistical scatter, and values of our LOWESS fit.}
    \centering
    \fontsize{8}{6.3}\selectfont 
    \begin{tabular}{c|c|c|c|c|c|c|c|c|c|c|c|c|c|c|c|c}
\hline\hline
log $M_\star$ 
& \multicolumn{2}{c|}{$1<z<1.5$} 
& \multicolumn{2}{c|}{$1.5<z<2$} 
& \multicolumn{2}{c|}{$2<z<2.5$} 
& \multicolumn{2}{c|}{$2.5<z<3$} 
& \multicolumn{2}{c|}{$3<z<4$} 
& \multicolumn{2}{c|}{$4<z<5$} 
& \multicolumn{2}{c|}{$5<z<7$} 
& \multicolumn{2}{c}{$7<z<9$} \\

{[$M_\odot$]} 
& \multicolumn{2}{c|}{} 
& \multicolumn{2}{c|}{} 
& \multicolumn{2}{c|}{} 
& \multicolumn{2}{c|}{} 
& \multicolumn{2}{c|}{} 
& \multicolumn{2}{c|}{} 
& \multicolumn{2}{c|}{} 
& \multicolumn{2}{c}{} \\
\hline\hline
          7.6&$-1.14\pm0.46$&$-1.23$ & & & & & & & & & & & & &\\
          7.8&$-1.08\pm0.47$&$-1.06$ &$-0.82\pm0.37$&$-0.89$& & & & & & & & & & &\\ 
          8.0&$-0.94\pm0.47$&$-0.90$&$-0.72\pm0.40$&$-0.74$&$-0.56\pm0.38$&$-0.60$& & & & & & & & &\\
          8.2&$-0.75\pm0.48$&$-0.73$&$-0.60\pm0.42$&$-0.57$&$-0.42\pm0.40$&$-0.42$&$-0.21\pm0.38$ &$-0.24$&$-0.18\pm0.28$&$-0.12$& & & & &\\
          8.4&$-0.57\pm0.48$&$-0.56$&$-0.46\pm0.43$&$-0.40$&$-0.27\pm0.41$&$-0.25$& $-0.03\pm0.38$&$-0.09$&$-0.03\pm0.32$&$-0.03$&$0.12\pm0.17$ &0.16 & & & &\\
          8.6&$-0.35\pm0.49$&$-0.40$&$-0.27\pm0.47$&$-0.24$&$-0.10\pm0.44$&$-0.08$&$0.07\pm0.39$& $0.06$&$0.14\pm0.32$& 0.12& $0.26\pm0.19$&0.28 &$0.29\pm0.17$ & 0.32&&\\
          8.8&$-0.18\pm0.47$&$-0.23$&$-0.09\pm0.48$&$-0.07$&$0.13\pm0.46$&$0.09$&$0.25\pm0.39$& 0.20&$0.27\pm0.31$&0.26 &$0.42\pm0.21$ &0.42 &$0.48\pm0.19$ &0.47 &$0.41\pm0.15$&0.41\\
          9.0&$-0.04\pm0.45$&$-0.06$&$0.14\pm0.47$&$0.09$&$0.35\pm0.45$&0.26&$0.40\pm0.38$&0.35&$0.45\pm0.32$ & 0.41& $0.55\pm0.22$&0.56 &$0.65\pm0.23$ &0.63 &$0.56\pm0.15$&0.63\\
          9.2&$0.16\pm0.43$&$0.10$&$0.31\pm0.47$&0.26&$0.52\pm0.46$&0.43&$0.57\pm0.41$&0.50&$0.58\pm0.32$ & 0.55& $0.70\pm0.23$&0.70 &$0.81\pm0.24$ &0.79 &$0.81\pm0.20$&0.86\\
          9.4&$0.29\pm0.40$&$0.26$&$0.47\pm0.44$&0.43&$0.57\pm0.48$&0.59&$0.68\pm0.44$&0.64&$0.71\pm0.34$ & 0.69&$0.83\pm0.24$ &0.85 &$0.94\pm0.26$ &0.95 &$1.19\pm0.30$&1.09\\
          9.6&$0.42\pm0.37$&$0.42$&$0.59\pm0.47$&0.58&$0.69\pm0.49$&0.74&$0.83\pm0.47$&0.78&$0.87\pm0.35$ & 0.83&$0.95\pm0.25$ &0.99 &$1.09\pm0.27$ &1.11 &$1.15\pm0.16$&1.31\\
          9.8&$0.57\pm0.35$&$0.57$&$0.74\pm0.46$&0.73&$0.85\pm0.47$&0.87&$0.89\pm0.45$&0.90& $1.00\pm0.36$& 0.96&$1.13\pm0.25$ &1.13 & $1.35\pm0.30$&1.27 &$1.52\pm0.09$&1.52\\
          10.0&$0.72\pm0.36$&$0.72$&$0.89\pm0.42$&0.88&$0.87\pm0.49$&1.01&$0.95\pm0.47$&1.02& $1.04\pm0.37$&1.09 &$1.30\pm0.26$ & 1.28&$1.32\pm0.35$ & 1.42&$1.89\pm0.36$&1.74\\
          10.2&$0.87\pm0.32$&$0.87$&$0.90\pm0.40$&1.03&$0.99\pm0.44$&1.14&$1.04\pm0.44$&1.15&$1.12\pm0.40$ & 1.22&$1.45\pm0.26$ & 1.43& $1.53\pm0.37$& 1.58&&\\
          10.4&$1.00\pm0.31$&$1.02$&$1.03\pm0.39$&1.18&$1.05\pm0.41$&1.28&$1.16\pm0.41$&1.27&$1.19\pm0.36$ & 1.35&$1.69\pm0.26$ &1.57 & $1.87\pm0.35$& 1.74&&\\
          10.6&$1.02\pm0.31$&$1.17$&$1.13\pm0.36$&1.32&$1.15\pm0.36$&1.41&$1.35\pm0.40$&1.39&$1.45\pm0.37$ & 1.48&$1.83\pm0.30$ &1.72 &$1.86\pm0.23$ & 1.91&&\\
          10.8&$1.19\pm0.29$&$1.32$&$1.22\pm0.31$&1.47&$1.29\pm0.32$&1.54&$1.37\pm0.33$&1.51&$1.67\pm0.27$&1.60 &$2.22\pm0.35$ &1.87 & $2.49\pm0.28$& 2.07&&\\
          
          11.0&$1.26\pm0.23$&$1.47$ &$1.31\pm0.28$&1.61&$1.41\pm0.26$&1.67& $1.47\pm0.39$&1.63&$1.83\pm0.39$& 1.73& $2.23\pm0.22$& 2.02&$2.40\pm0.22$ &2.24 &&\\
          \hline
    \end{tabular}
    \label{tab:running_medians}
\tablefoot{For each redshift interval, the first column shows the running medians and scatter at that stellar mass. The second shows the value of the LOWESS fit. We only include the values down to our mass-completeness limits.}
\end{table*}

\section{Main sequence tests}
\label{app:MS_tests}

In Sect.~\ref{sec:results}, we report values of the MS scatter that decrease with increasing redshift. Although these estimates lie within the canonical values, in this appendix we explore the possible underlying causes for this trend. We also show a fit based only on JWST data, excluding CANDELS, SHARDS/CANDELS FAINT, and COSMOS2020.

\subsection{S/N test}

We selected those galaxies with a S/N~$>50$ in five or more bands. This is a very restrictive cut, which allowed us to select only those galaxies with very robust individual measurements from \texttt{Dense Basis}. We repeated the analysis described in Sect.~\ref{sec:results}, weighting by the survey area as detailed in Sect.~\ref{sec:properties}, and show the corresponding MS fits in Fig.~\ref{fig:MS_app}. The values of the slopes and scatter are included in Table \ref{tab:slopes_app}.

\begin{table}[H]
\caption{Evolution of the Main Sequence slopes and scatter with redshift in the global and low-$M_\star$ regimes that result when applying an S/N cut.}
\setlength{\tabcolsep}{2.4pt}
    \centering
    \small
    \begin{tabular}{c|c|c|c|c|c}
      $z$ bin   &  & \multicolumn{2}{c|}{Instrinsic - symm} & \multicolumn{2}{c|}{Intrinsic - asymm} \\ \hline
      & &$\beta$&$\epsilon$&$\beta$&$\epsilon$\\
      \hline
      $1\leq z \leq 1.5$ & All galaxies & 0.68 & 0.36& 0.64 & 0.38\\
         & Low-$M_\star$ & 0.21 & 0.37 & 0.23 & 0.37 \\
         \hline
      $1.5\leq z \leq 2$ & All galaxies & 0.61 & 0.37& 0.58 & 0.38\\
         & Low-$M_\star$ & 0.16 & 0.36 & 0.27 & 0.36\\
         \hline
    $2\leq z \leq 2.5$ & All galaxies & 0.57 & 0.36& 0.57 & 0.36\\
         & Low-$M_\star$ & 0.40 & 0.35 & 0.48 & 0.34\\
         \hline
    $2.5\leq z \leq 3$ & All galaxies & 0.56 & 0.40& 0.55 & 0.40\\
         & Low-$M_\star$ & 0.25 & 0.36 & 0.31 & 0.37\\
         \hline
    $3\leq z \leq 4$ & All galaxies & 0.58 & 0.33& 0.59 & 0.34\\
         & Low-$M_\star$ & 0.47 & 0.32 & 0.50 & 0.33\\
         \hline
    $4\leq z \leq 5$ & All galaxies & 0.77 & 0.24& 0.77 & 0.25\\
         & Low-$M_\star$ & 0.71 & 0.22 & 0.68 & 0.23\\
         \hline
        $5\leq z \leq 7$ & All galaxies & 0.88 & 0.17& 0.91 & 0.21\\
         & Low-$M_\star$ & 0.95 & 0.16 & 0.88 & 0.18\\
         \hline
    $7\leq z \leq 9$ & All galaxies & 1.07 & 0.15& 1.08 & 0.15\\
         & Low-$M_\star$ & 1.37 & 0.12 & 1.19 & 0.13\\
         \hline
    \end{tabular}
    \label{tab:slopes_app}
    \tablefoot{For this test, we focused on the intrinsic MS parameters, computed using \texttt{lmfit} and \texttt{emcee}.}
\end{table}

Looking at Fig.~\ref{fig:MS_app}, we see how the low-mass galaxies that pass this screening are those with the highest SFRs. This translates into a flattening of the MS (yellow line), which reflects a selection effect, even though the global sample is mass complete at those $M_\star$. The latter highlights the need to provide accurate and realistic estimates for completeness in this kind of study, based on the final screened samples (see also \citealt{Simmonds2025}).

If we now focus on Table~\ref{tab:slopes_app} and the values retrieved for the scatter for all the galaxies, we can see that these are consistent with 0.3 dex. They do not evolve with $z$ up to $z\sim5$, above which this S/N cut resembles the different screening processes applied in Sect.~\ref {sec:data_sample} and \ref{sec:properties}, ending up with a similar selection.

\subsection{JWST data only}

\begin{figure*}[htp]
    \centering
\includegraphics[width=\textwidth, height=17cm]{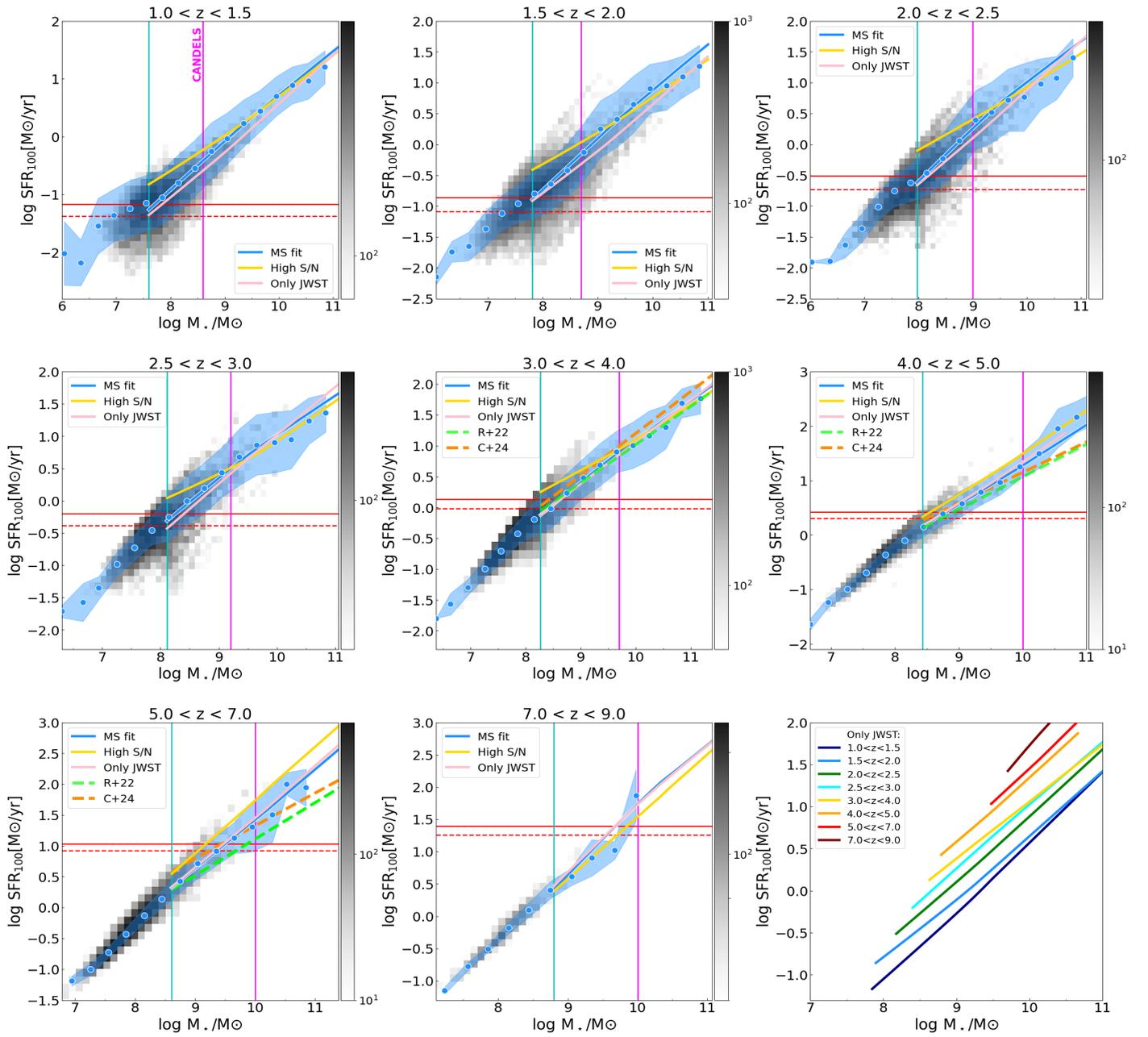}
    \caption{Star formation rates versus stellar masses divided into different redshift intervals covering ${1<z<9}$. The 2D histograms represented here correspond only to the JWST data. The LOWESS fit to the main sequence described in Sect.~\ref{sec:results} is shown in blue; in yellow, we include the MS fit using only those galaxies with S/N~$>50$ in at least five bands; in pink, we show the fits derived using only JWST data. We include two fits to the MS from the literature as dashed lines: lime for \citealt{Rinaldi2022} and orange for \citealt{Clarke2024}.}
    
    \label{fig:MS_app}
\end{figure*}

In this appendix we include the MS fits derived using only JWST observations, namely data from CEERS, JADES, CANUCS, and UNCOVER. These fits are shown in Fig.~\ref{fig:MS_app} (pink line), where we also depict the fits from \citet{Rinaldi2022} and \citet{Clarke2024}, which probe $z>3$ galaxies. 

Examining the evolution of the fits based just on JWST data at $z<3$, we see that they lie below the MS fits computed using the whole sample. The final panel includes all these fits, and we can see how the slopes, especially at ${1\leq z \leq 2}$, get flatter in the low-$M_\star$ regime, contrary to what we find using the whole sample. This illustrates the challenges shown in Fig.~\ref{fig:ms_drawing} in fitting the MS when lacking a comprehensive $M_\star$ and SFR coverage, and thus in understanding the evolution of the MS parameters. At higher $z$, the contribution from JWST surveys to the number counts is larger, resulting in similar results to those described in Sect.~\ref{sec:results}.

Examining the evolution of the fits from \citet{Rinaldi2022} and \citet{Clarke2024}, we see that their slopes become increasingly flatter. At $z>5$, the fits fall below our SFR and $M_\star$ limits, pointing to potential selection effects.

\end{document}